\pdfoutput=1
\documentclass[aps,showpacs,showkeys,floatfix]{revtex4}
\usepackage[pdftex]{graphicx}\usepackage{graphicx}
\usepackage{amssymb}
\usepackage{amsmath}
\usepackage{mathptmx} 
\usepackage{latexsym}
\usepackage{hyperref}
\begin{document}

\title{Magnetic susceptibility of topological semimetals}

\author{G.~P.~Mikitik}
\affiliation{B.~Verkin Institute for Low Temperature Physics \&
Engineering, Ukrainian Academy of Sciences, Kharkiv 61103,
Ukraine}

\author{Yu.~V.~Sharlai}
\affiliation{B.~Verkin Institute for Low Temperature Physics \&
Engineering, Ukrainian Academy of Sciences, Kharkiv 61103,
Ukraine}

\begin{abstract}
We give a review of theoretical and experimental results concerning the magnetic susceptibility of the Weyl, Dirac, and nodal-line  semimetals. In particular, dependences of the susceptibility on the chemical potential, temperature, and magnitude of the magnetic field are discussed. The presented results show that the specific features of the magnetic susceptibility can serve as a hallmark of the topological semimetals, and hence magnetic measurements can be useful in investigating these materials.
\end{abstract}


\maketitle

\section{Introduction}
\label{intro}

In recent years much attention has been given to the topological Weyl, Dirac, and nodal-line semimetals; see, e.g., recent reviews \cite{armit,bernevig,gao,wang-r,weng-r,fang-r,chiu} and references therein. In the Weyl semimetals, the electron bands contact at discrete (Weyl) points of the Brillouin zone and disperse linearly in all directions around these critical points. The same type of the band contact occurs in the Dirac semimetals, but the bands are double degenerate in spin, i.e., a Dirac point can be considered as a superposition of two Weyl points in the quasi-momentum space. The chemical potential of electrons in the Weyl and Dirac semimetals is implied to be close to the band-contact energy $\varepsilon_d$.
In the nodal-line semimetals the conduction and valence bands touch along lines in the Brillouin zone and disperse linearly in directions perpendicular to these lines. It is necessary to emphasize that the contact of the electron energy bands along the lines is the widespread phenomenon in crystals \cite{herring,m-sh14,kim,fang}. For example, such contacts of the bands occur in graphite \cite{graphite}, beryllium \cite{beryl,beryl1}, magnesium \cite{beryl1}, aluminium \cite{al}, LaRhIn$_5$ \cite{prl04}. However, the degeneracy energy of the bands, $\varepsilon_d$, is not constant along such lines, and the $\varepsilon_d$ varies in a finite interval between its minimum $\varepsilon_{min}$ and maximum $\varepsilon_{max}$ values, reaching them at certain points of the line. A crystal with the band-contact line can be named the topological semimetal if the difference $\varepsilon_{max}- \varepsilon_{min}\equiv 2\Delta$ is sufficiently small and if the chemical potential $\zeta$ of the electrons does not lie far away from the mean energy $\varepsilon_d^0 \equiv (\varepsilon_{max}+ \varepsilon_{min})/2$ of the line. A number of various Dirac \cite{wang,neupane,borisenko,jeon,liang,ali,liu,crassee,z.wang,liu1,gibson}, Weyl \cite{h.weng,huang,S-Xu,Lv,sol,c.wang,ruan,verg}, and nodal-line semimetals \cite{kim,fang,hei1,pie,weng,mullen,xie,yu,yama,phil,schoop,neupane1,schoop1,huang1,bian,chen} were predicted in recent years (see also \cite{zhang19}), and a part of these predictions have already been confirmed experimentally.

It is well known that the topological semimetals are characterized by  topologically protected surface states \cite{hei1,wan,balents}, the giant transverse magnetoresistance \cite{liang,he-r(h),shek,ali-wte2,lv-r(h)}, the nonlocal transport \cite{para,zhang-nonl}, the negative longitudinal magnetoresistance \cite{niel,son} related with the so-called chiral anomaly, the anomalous Hall effect \cite{burk}, and the quantum oscillations associated with  surface Fermi arcs \cite{pott,moll}.
In this paper we call attention to the specific features of the bulk   magnetic susceptibility of electrons in the topological semimetals. These features can be considered as one more hallmark of these materials.  The specific properties of the susceptibility for the topological  semimetals are due to the following: The Weyl and Dirac points, as well as the points of the band-contact lines with the energies $\varepsilon_{min}$ and $\varepsilon_{max}$, are the points of the electron topological transitions. In other words,
if the electron chemical potential $\zeta$, which can vary with doping and crystal deformations, passes through one of the critical energies $\varepsilon_d$, $\varepsilon_{min}$,  $\varepsilon_{max}$, the topology of the appropriate Fermi surface changes. However, these Fermi-surface transformations are not the $2\frac{1}{2}$-order topological transitions analyzed by Lifshitz \cite{lif}. These transitions have higher orders according to the classification of Ref.~\cite{lif}, and for weak magnetic fields,  $\hbar \omega_c< T$, they are accompanied with giant anomalies in the orbital part of the magnetic susceptibility \cite{m-sv}. Here $T$ is the temperature, and $\omega_c$ is the electron cyclotron frequency which determines the spacing $\Delta\varepsilon_H$ between the Landau subbands in the magnetic field, $\Delta\varepsilon_H=\hbar \omega_c$. At higher magnetic fields, $\hbar\omega_c>T$, the well-known de Haas - van Alphen oscillations of the magnetization appear \cite{Sh}. However, these  oscillations in the topological semimetals are shifted in phase as compared to the usual case discussed by Shoenberg \cite{Sh}. This shift is essentially associated with nonzero Berry phase generated by the Weyl and Dirac points \cite{berry} and by the band-contact lines \cite{prl}. With further increase of the magnetic field  when $\Delta\varepsilon_H$ exceeds not only $T$ but also $|\zeta-\varepsilon_d|$ or $|\zeta-\varepsilon_{max}|$ (or $|\zeta-\varepsilon_{min}|$), the magnetization in this ultra-quantum regime exhibits special dependences on the magnetic field $H$. The types of these dependences are determined by the electron dispersion relations in the vicinity of the Dirac (Weyl) points and of the nodal lines.

The paper is organized as follows: In Sec.~\ref{spectr}, we describe the electron spectra of the Weyl and Dirac semimetals without magnetic field and with applied $H$. In Sec.~\ref{d-w} these spectra are used in analyzing the electron magnetic susceptibility in the weak and strong magnetic fields. In particular, we consider the feature of de Haas - van Alphen (and Shubnikov - de Haas) oscillations in these semimetals. In Sec.~\ref{line}, the electron spectrum and the  magnetic susceptibility of the nodal-line semimetals are discussed. Conclusions are presented in Sec.~\ref{conc}. It should be also noted that throughout this review  article we restrict the theoretical analysis of the susceptibility to the case of noninteracting electrons. Although the electro-electron interaction in the Weyl (Dirac) and nodal-line  semimetals was considered, e.g., in Refs.~\cite{niss17,niss18,ghosh19,huh16,liu17,Roy17}, its effect on the bulk  magnetic susceptibility of these materials  remains to be studied.

\section{Electron spectrum in Dirac and Weyl semimetals} \label{spectr}

\subsection{Spectrum without magnetic field}\label{sec2.1}

The Dirac and Weyl semimetals are characterized by a sufficiently strong spin-orbit interaction. With this interaction the most general ${\bf k}\cdot{\bf p}$ Hamiltonian $\hat H$ for the conduction and valence electron bands in the vicinity of a {\it Dirac} point  has the form \cite{m-sv}:
 \begin{eqnarray}\label{1}
\hat H=\left (\begin{array}{cccc} {\mathcal E}_{c} & R & 0 & S \\ R^* & {\mathcal E}_{v} & -S & 0 \\ 0 & -S^* & {\mathcal E}_{c} & R^* \\ S^* & 0 & R & {\mathcal E}_{v} \\
\end{array} \right),
 \end{eqnarray}
where
 \begin{eqnarray}\label{2}
{\mathcal E}_{c,v}&=&\varepsilon_d+{\bf v}_{c,v}\cdot {\bf p}, \nonumber \\
R&=&{\bf r}\cdot {\bf p}, \\
S&=&{\bf s}\cdot {\bf p},  \nonumber
  \end{eqnarray}
the quasi-momentum ${\bf p}$ is measured from the Dirac point;  ${\bf v}_{c,v}$ are intraband and ${\bf r}$ and ${\bf s}$ are interband matrix elements of the velocity operator calculated at ${\bf p}=0$; the vectors ${\bf v}_{c,v}$ are real, while ${\bf r}$ and ${\bf s}$ are generally complex quantities. In Hamiltonian (\ref{1}) we have taken into account only the time-reversal symmetry and a twofold spin degeneracy of the electron bands in centrosymmetric crystals. In reality, the electron Hamiltonian for a topological semimetal is even simpler than that given by Eqs.~(\ref{1}), (\ref{2}) \cite{yang-cl,gao-cl} since for the Dirac point to be stable,
an additional crystalline symmetry (other than the time-reversal and inversion symmetries) is necessary \cite{young}; see, e.g., the electron spectrum of Na$_3$Bi below.

Diagonalization of the  Hamiltonian (\ref{1}), (\ref{2}) gives  the   dispersion  relations for  the  electron bands in the   vicinity of  the Dirac point:
 \begin{eqnarray}\label{3}
 \varepsilon_{c,v}({\bf p})&=&\varepsilon_d+{\bf a}\cdot{\bf p}+E_{c,v}({\bf p}), \\
 E_{c,v}({\bf p})&=&\pm\{({\bf a'}\cdot {\bf p})^2+|R|^2+|S|^2 \}^{1/2}, \label{4}
 \end{eqnarray}
where  the   following notations have been introduced:
 \begin{eqnarray*}
 {\bf a}=({\bf v}_c+{\bf v}_v)/2;\ \ \ {\bf a'}=({\bf v}_c-{\bf v}_v)/2.
 \end{eqnarray*}
Equation (\ref{4}) shows that $E_{c,v}^2$ is a quadratic form in the components of the vector ${\bf p}$. Hereafter we choose the coordinate axes along principal directions of this form. Let  $b_{ii}$ ($i=1, 2, 3$)  be its principal values, i.e.,
 \begin{eqnarray}\label{5}
E_{c,v}^2=b_{11}p_1^2+b_{22}p_2^2+b_{33}p_3^2,
 \end{eqnarray}
where $b_{ii}$ are expressible in terms of  the   components of  the   vectors  ${\bf a'}$, ${\bf r}$,  ${\bf s}$.
The scaling of coordinate axes, $\tilde p_i=p_i\sqrt{b_{ii}}$, transforms Eqs.~(\ref{3}), (\ref{5}) into the form that depends on the constant dimensionless vector $\tilde {\bf a}$ only:
 \begin{eqnarray*}
 \varepsilon_{c,v}&=&\varepsilon_d+\tilde{\bf a}\cdot\tilde{\bf p}\pm |\tilde {\bf p}|,
 \end{eqnarray*}
where its components are defined by $\tilde a_i=a_i/\sqrt{b_{ii}}$. The vector $\tilde{\bf a}$ characterizes a tilt of the spectrum. When  the length of $\tilde {\bf a}$ is less than unity,
\begin{eqnarray*}
\tilde a^2=\frac{a_1^2}{b_{11}}+ \frac{a_2^2}{b_{22}} +\frac{a_3^2}{b_{33}} < 1,
 \end{eqnarray*}
the dispersion relations $\varepsilon_{c,v}({\bf p})$ looks like in Fig.~1a. In this case, at $\zeta<\varepsilon_d$, the Fermi surface is a {\it hole} closed pocket which, at $\zeta=\varepsilon_d$, shrinks into the point ${\bf p}=0$, and at $\zeta > \varepsilon_d$ a new closed {\it electron} Fermi pocked appears, Fig.~1a. Since near $\varepsilon_d$ a singular part of the electron energy of the crystal is proportional to $(\zeta-\varepsilon_d)|\zeta-\varepsilon_d|^3$, the electron topological transition at $\zeta=\varepsilon_d$ can be named as the $4$-order transition according to the classification of Lifshitz \cite{lif}. When $\tilde a^2>1$, there is always a direction in the ${\bf p}$-space along which the dispersion relations $\varepsilon_{c,v}({\bf p})$ look like in Fig.~1b, and open electron and hole pockets of the Fermi surface exist both at $\zeta < \varepsilon_d$ and $\zeta > \varepsilon_d$. Thus, there is no topological transition with changing $\zeta$ in the case $\tilde a^2>1$.

\begin{figure}[tbp] 
 \centering  \vspace{+9 pt}
\includegraphics[scale=.90]{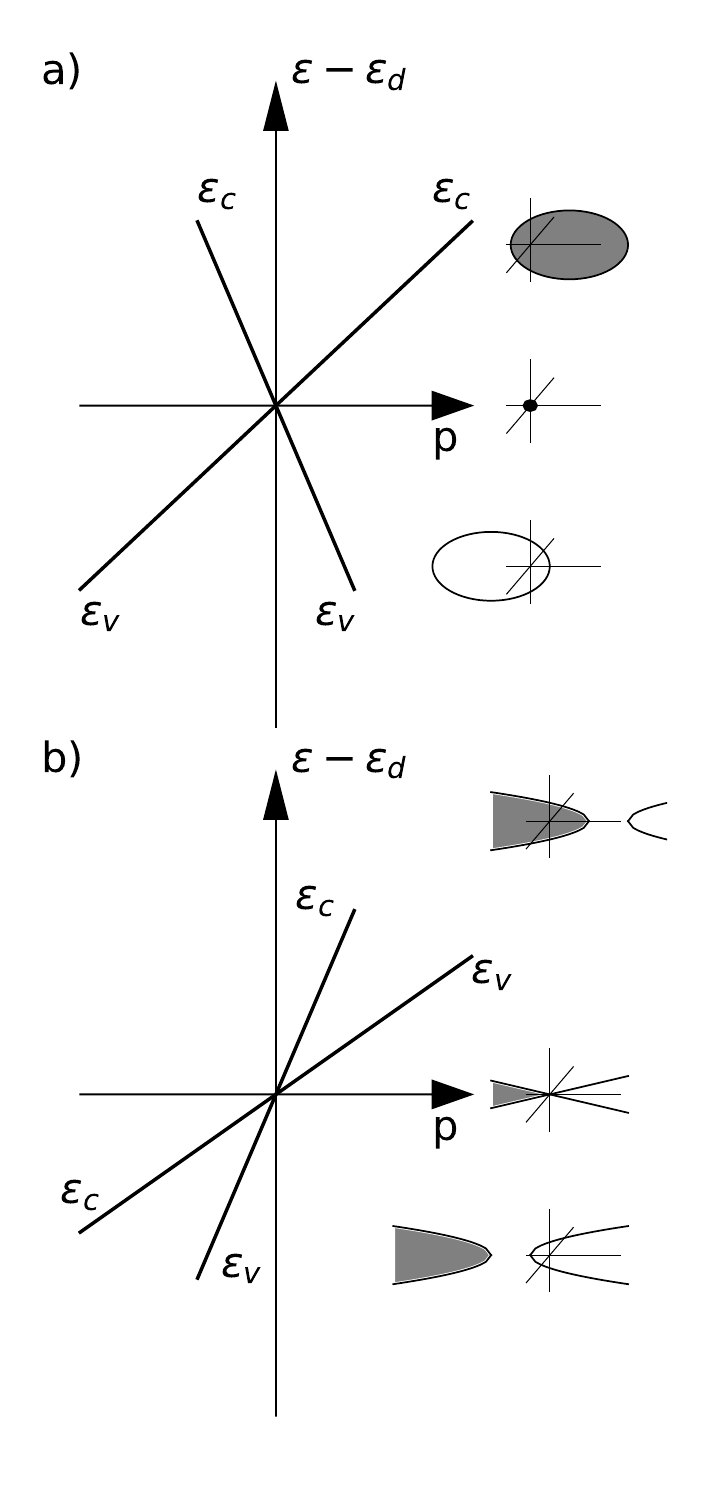}
\caption{\label{fig1} Dispersion relations $\varepsilon_c(p)$ and $\varepsilon_v(p)$ of the electron energy bands in the vicinity of their degeneracy point in the cases of $\tilde a^2<1$ (a) and $\tilde a^2>1$ (b). On the right the Fermi surfaces at $\zeta-\varepsilon_d<0$, $\zeta-\varepsilon_d=0$, and $\zeta-\varepsilon_d>0$ are shown together with the degeneracy point which is the origin of the coordinate axes. The shaded and white surfaces correspond to the electron and hole charge carriers, respectively.
 } \end{figure}   

The most general ${\bf k}\cdot{\bf p}$
Hamiltonian $\hat H$ for the conduction and valence electron bands in the vicinity of a {\it Weyl} point has the form:
 \begin{eqnarray}\label{6}
\hat H= (\varepsilon_d + {\bf a}\cdot{\bf p})\sigma_0 + ({\bf a}'\cdot {\bf p})\sigma_z + ({\bf v}^{(1)}\cdot {\bf p})\sigma_x +({\bf v}^{(2)}\cdot {\bf p})\sigma_y .
 \end{eqnarray}
where ${\bf a}$, ${\bf a}'$, ${\bf v}^{(1)}$, ${\bf v}^{(2)}$ are real constant vectors, $\sigma_0$ is the unit matrix, and $\sigma_i$ are the Pauli matrices. In fact, this Hamiltonian coincides with the upper $2\times 2$ block of Hamiltonian (\ref{1}) if one defines the complex vector ${\bf r}$ in Eq.~(\ref{2}) as follows: ${\bf r}={\bf v}^{(1)}- i{\bf v}^{(2)}$. Thus, formulas (\ref{3})-(\ref{5}) and the discussion accompanying them refer equally not only to the Dirac points but also to the case of the Weyl points.

It is necessary to emphasize that the parameter $\tilde a^2$ specifying the tilt of the spectrum generally differs from zero for the Dirac and Weyl points. For example, such tilted spectra occur in Na$_3$Bi, Cd$_3$As$_2$, WTe$_2$, and in the TaAs family of the  semimetals. If $\tilde a^2<1$, one has a type-I Weyl (Dirac) semimetal, while the case $\tilde a^2>1$ corresponds to the so-called type-II Weyl (Dirac) semimetals \cite{sol}. In particular, the spectrum of Na$_3$Bi in the vicinity of any of its two Dirac points is described by Eqs.~(\ref{3}), (\ref{5}) with ${\bf a}=(0,0,a_3)$, $a_3=\pm 2C_1p_z^c/\hbar^2$, $b_{11}=b_{22}=A^2$, $b_{33}=(2M_1p_z^c/\hbar^2)^2$ where $A$, $C_1$, $M_1$, $p_z^c$ are the constant parameters introduced in Refs.~\cite{z.wang}, the axes $1$, $2$, and $3$ coincide with the crystallographic axes $x$, $y$, $z$, and the two Dirac points lie in the $z$-axis with the coordinates $p_z=\pm p_z^c$. Then, using the data of Ref.~\cite{z.wang}, we find $\tilde a^2=(C_1/M_1)^2\approx 0.68$. In other words, Na$_3$Bi is the type-I Dirac semimetal with the tilted
spectrum. At small $|\zeta-\varepsilon_d|\neq 0$, its Fermi surface near any of the two Dirac points is an ellipsoid, with the center of the ellipsoid being shifted relative to the point $p_z=\pm p_z^c$ by the value $\Delta p_z=-a_3(\zeta-\varepsilon_d)/(b_{33}-a_3^2)$. Note that the shift of this sort occurs for any Dirac (Weyl) point with the tilted spectrum, Fig.~\ref{fig1}.

Curiously, investigations of the type-I and type-II Weyl (Dirac) points can shed light on the astrophysical problems \cite{vol16,vol17}.

\subsection{Spectrum in the magnetic field}\label{sec2.2}

In the case of Hamiltonian (\ref{1}), (\ref{2}), the appropriate electron spectrum in the magnetic field $H$ was obtained at an arbitrary direction of ${\bf H}$ and at any value of the parameter $\tilde a^2$ many years ago  \cite{m-sh}. The derived formulas immediately describe the electron spectrum in the magnetic field near the Dirac points. Similar formulas for the case of the Weyl points were independently obtained in the papers \cite{yu16,udag,tch}. We present here the results of Ref.~\cite{m-sh} and then point out a small difference between the spectra for the Dirac and Weyl points.

In the magnetic field ${\bf H}={\bf n}H$ directed along a unit vector ${\bf n}$, the Landau subbands $\varepsilon^l(p_n)$ of a Dirac point are found from the equation \cite{m-sh}:
\begin{equation}\label{7}
S(\varepsilon^{l},p_n)=\frac{2\pi\hbar e H}{c}l,
\end{equation}
where $e$ is the absolute value of the electron charge,  $l=0,1,2,\dots$, and $S(\varepsilon^{l},p_n)$ is the area of the cross-section of the constant-energy surface $\varepsilon_{c,v}({\bf p})=\varepsilon^l$ by the plane $p_n=$ const. Here $p_n= {\bf p}\cdot {\bf n}$ is the component of the quasi-momentum along the magnetic field. Although Eq.~(\ref{7}) looks like the semiclassical quantization condition, this formula determines the {\it exact} spectrum, i.e., it gives correct $\varepsilon^{l}(p_n)$ even at $l\sim 1$. We shall show in Sec.~\ref{sec2.3} that the exact and semiclassical spectra in the magnetic fields really coincide at all $l$ for the electrons with dispersion relation described by Eqs.~(\ref{3}) and (\ref{5}). In the explicit form Eq.~(\ref{7}) yields
 \begin{eqnarray}\label{8}
 \varepsilon_{c,v}^l(p_n)&=&\varepsilon_d +vp_n \pm \left[\frac{e\hbar \alpha H}{c}\,l+L\cdot (p_n)^2 \right]^{1/2},
 \end{eqnarray}
where
\begin{eqnarray}\label{9}
\alpha&=&\frac{2R_n^{3/2}}{b_{11}b_{22}b_{33}\tilde{\bf n}^2}, \nonumber \\
L&=&\frac{R_n}{b_{11}b_{22}b_{33}\tilde{\bf n}^4}=\frac{(1-\tilde a^2)\tilde{\bf n}^2+(\tilde {\bf a}\cdot\tilde{\bf n})^2}{\tilde{\bf n}^4}, \nonumber \\
R_n\!\!&=&\sum_{i,j=1}^{3}\kappa^{ij}n_in_j= b_{11}b_{22}b_{33}[(1-\tilde a^2)\tilde{\bf n}^2+(\tilde {\bf a}\cdot \tilde{\bf n})^2], \\
\kappa^{ij}\!\!&=&\frac{b_{11}b_{22}b_{33}}{(b_{ii}b_{jj})^{1/2}} \left[(1-\tilde a^2)\delta_{ij}+\tilde a_i\tilde a_j\right],\nonumber \\
v&=&\frac{(\tilde {\bf a}\cdot \tilde{\bf n})}{\tilde{\bf n}^2}, \nonumber
 \end{eqnarray}
the components of the vector $\tilde{\bf n}$ are determined by the relation: $\tilde n_i=n_i/\sqrt{b_{ii}}$, and hence
\[
\tilde{\bf n}^2=\frac{n_1^2}{b_{11}}+\frac{n_2^2}{b_{22}}+\frac{n_3^2}{b_{33}}.
\]
All the Landau subbands $\varepsilon_{c,v}^l(p_n)$ of the Dirac point  are double degenerate in spin apart from the subbands $\varepsilon_{c}^0(p_n)$  and $\varepsilon_{v}^0(p_n)$ which are nondegenerate. In deriving Eqs.~(\ref{7})-(\ref{9}) we have neglected the direct interaction $(e\hbar/mc){\bf s}\cdot{\bf H}$ of the electron spin ${\bf s}$ with the magnetic field ${\bf H}$. The impact of this interaction on the Landau subbands (and on the magnetic susceptibility) is relatively small and is of the order of $m_*/m$ where $m$ is the electron mass and $m_*$ is its cyclotron mass, see below.

For given ${\bf n}$, spectrum (\ref{8}) describing the Landau subbands $\varepsilon^l_{c,v}(p_n)$ exists only if $R_n>0$. The geometrical meaning of $R_n$ is the following: The quantity $R_n$ is positive when the boundary of the cross-section of the  Fermi surface by the plane perpendicular to $p_n$ is a closed curve (an ellipse), and hence the appropriate cross-section area is finite. When $\tilde a^2<1$, the $R_n$ is positive at {\it any} direction of the magnetic field. This is also evident from the Fermi surfaces shown Fig.~1a. When $\tilde a^2>1$, there are directions of the magnetic field for which $R_n<0$ and spectrum (\ref{8}) does not exist; see Fig.~1b. In other words, for the type-II semimetals, a rotation of the magnetic field leads to collapse of the Landau subbands in a certain region of its directions.

The electron spectrum in the magnetic field for the case of the Weyl points is described by the same formulas (\ref{7})-(\ref{9})  \cite{m-sh,yu16,udag,tch}, but the Landau subbands $\varepsilon_{c,v}^l(p_n)$ are nondegenerate, and the subband $\varepsilon^0(p_n)$ corresponding to $l=0$ is shared between the branches ``$c$'' and ``$v$''. This subband has the form \cite{yu16,udag,tch}:
\begin{eqnarray}\label{10}
 \varepsilon^0(p_n)&=&\varepsilon_d + (v - q_{ch}\sqrt L)p_n,
 \end{eqnarray}
where $q_{ch}={\rm sign}({\bf a}'\cdot[{\bf v}^{(1)}\times{\bf v}^{(2)}])=\pm 1$ is the chirality \cite{armit} of the Weyl point defined by Hamiltonian (\ref{6}). It is also worth noting that the $p_n$-dependence of the zeroth Landau subband describing by    Eq.~(\ref{8}) or Eq.~(\ref{10}) generally differs from the dispersion of the bands $\varepsilon_{c,v}({\bf p})$ in the direction ${\bf n}$ of the magnetic field if ${\bf a}\neq 0$ in Eqs.~(\ref{3}) and (\ref{6}). This difference is due to the shift of the center of the ellipsoid $\varepsilon_{c,v}({\bf p})=\zeta$ relative to the Dirac (Weyl) point  with the tilted spectrum, see Sec.~\ref{sec2.1} and Fig.~\ref{fig1}.

Finally, it should be mentioned that the electron spectrum in the magnetic field can be found for Hamiltonian (\ref{1}), (\ref{2}) in which an energy gap between the bands ``$c$'' and ``$v$'' is introduced, and this spectrum is still described by Eq.~(\ref{7}) \cite{m-sh}.

\subsection{Semiclassical quantization} \label{sec2.3}

It is instructive to compare the semiclassical electron spectrum in the magnetic field near the Dirac points with the exact one presented in Sec.~\ref{sec2.2}. If the electron bands in a crystal are double degenerate in spin, the general semiclassical quantization condition has the form \cite{Sh}:
\begin{equation}\label{11}
S(\varepsilon^{l},p_n)=\frac{2\pi\hbar e H}{c}\left(l+\frac{1}{2}\pm \frac{gm_*}{4m}\right),
\end{equation}
where $g$ is the electron $g$ factor; $m$ and $m_*=(1/2\pi)\partial S(\varepsilon,p_n)/\partial\varepsilon$ are the electron and cyclotron masses, respectively, and the other quantities are defined in the text near formula (\ref{7}). The theory of the $g$ factor for itinerant electrons was elaborated in Refs.~\cite{jetp,g1}. This theory based on the ideas of Roth \cite{roth} takes into account the electron-spin dynamics caused by the spin-orbit interaction when the electron moves in its orbit in the magnetic field. Using this theory (and neglecting the Zeeman term $e\hbar{\bf s}\cdot{\bf H}/mc$), it was shown \cite{g2} that $g=2m/m_*$ for any electron orbit in the case of Hamiltonian (\ref{1}), (\ref{2}). Insertion of this value of the $g$ factor into formula (\ref{11}) reproduces equation (\ref{7}) determining the exact electron spectrum in the magnetic field. Interestingly, this large value $2m/m_*$ of the $g$ factor would  occur even if the spin-orbit interaction were weak. In this limiting case the $g$ factor is the sum of the two terms, $g=g_1+g_2=2m/m_*$, where the first term $g_1$ is determined by the Berry phase $\Phi_B$ of the electron orbit while the second term $g_2$ is specified by an interband part $L$ of the electron orbital moment. If one considers a semiclassical electron as a wave packet, this $L$ can be interpreted as the orbital moment associated with self-rotation of the wave packet around its center of mass \cite{niu}. For the strong spin-orbit interaction the $g$ factor generally is not decomposed into the two independent parts determined by the Berry phase and the momentum, respectively. It is also worth noting that the result $g=2m/m_*$ remains true even if a gap appears in the Dirac spectrum \cite{g2}.

In the case of a Weyl point the electron states are nondegenerate in spin, and the general semiclassical quantization condition can be written as follows  \cite{Sh}:
\begin{equation}\label{12}
S(\varepsilon^{l},p_n)=\frac{2\pi\hbar e H}{c}\left(l+\gamma\right),
\end{equation}
where $\gamma$ is a constant. This constant depends on the Berry phase $\Phi_B$ of the electron orbit and on interband part $L$ of the electron orbital moment \cite{fuchs,m-sh12,alex}:
\begin{equation} \label{13}
\gamma -\frac{1}{2}=-\frac{\Phi_B}{2 \pi} -\frac{1}{2 \pi\hbar m } \oint \frac{L({\bf p})} {v_{\perp}({\bf p})} d{\bf p} ,
\end{equation}
where $v_{\perp}$ is the absolute value of projection of the electron velocity ${\bf v}= \partial \varepsilon_{c,v}({\bf p})/\partial {\bf
p}$  on the plane perpendicular to ${\bf H}$, and the integration is carried out over the electron orbit in the Brillouin zone. Since for a nondegenerate band the direction of electron spin is uniquely dictated by an appropriate point of the ${\bf p}$-space,
an electron is not free to adjust its spin to the orbital motion as it occurs in the double degenerate bands. It is for this reason that at any strength of the spin-orbit interaction, the difference $\gamma-\frac{1}{2}$ is the sum of the two independent terms associated with the Berry phase $\Phi_B$ and with the orbital moment $L$ averaged over the trajectory.

Fuchs et al. \cite{fuchs} calculated the right hand side of Eq.~(\ref{13}) for the two-dimensional electrons with the  Hamiltonian:
 \begin{eqnarray}\label{14}
\hat H= \Delta\cdot \sigma_z + {\rm Re}(f(p_x,p_y))\cdot \sigma_x -{\rm Im}(f(p_x,p_y))\cdot \sigma_y ,
 \end{eqnarray}
and obtained the universal value $\gamma -\frac{1}{2}=-\frac{1}{2}$.  In Eq.~(\ref{14}), $\Delta$ is a constant, and $f(p_x,p_y)$ is a complex function of $p_x$ and $p_y$. It was shown in Ref.~\cite{alex} (Appendix C) that for any electron orbit in the vicinity of a Weyl point, Hamiltonian (\ref{6}) can be reduced to the form (\ref{14}) by an appropriate unitary transformation if ${\bf a}=0$. This means that at least for the Weyl points without the tilt of their spectra, one has $\gamma=0$, and the semiclassical spectrum in the magnetic field coincides with the exact one. However, the semiclassical form of Eq.~(\ref{7}) seems to suggest that this statement remains true in the general case ${\bf a}\neq 0$. Note that in neglect of spin and the spin-orbit interaction, the constant $\gamma$ for electrons in an isolated band separated from other bands of the crystal by energy gaps takes another universal value $\gamma=1/2$ since both $\Phi_B$ and $L$ vanish.

\section{Magnetic susceptibility of Dirac and Weyl semimetals}\label{d-w}

When the chemical potential $\zeta$ of electrons in a Dirac (Weyl) semimetal lies near the degeneracy energy $\varepsilon_d$, the total magnetization ${\bf M}_{\rm tot}$ and the total magnetic susceptibility tensor $\chi^{ij}_{\rm tot}$ consist of their special parts ${\bf M}$ and $\chi^{ij}$ determined by the electron states located near the Dirac (Weyl) point and the background terms ${\bf M}_0$ and $\chi^{ij}_0$ specified by electron states located far away from this point,
\begin{eqnarray*}
M_{\rm tot}^{i}&=&M^{i}+M_{0}^{i}=M^i+ \sum_{j=1}^3\chi^{ij}_0H_j,\\  \chi_{\rm tot}^{ij}&=&\chi^{ij} +\chi^{ij}_0.
\end{eqnarray*}
It is the special term $\chi^{ij}$ that is responsible for dependences of the susceptibility on the chemical potential, temperature, and magnitude of the magnetic field. The background part $\chi^{ij}_0$ is practically constant and has no effect on the above-mentioned dependences. This part can have an impact only on angular dependence of ${\bf M}_{\rm tot}$ because the background magnetization ${\bf M}_0$ generally is not isotropic. Below we do not consider the background terms and discuss only the special parts of the magnetization and/or of the magnetic susceptibility which is assumed to be normalized to the unit volume.

The magnetic-field behavior of the susceptibility essentially depends on interrelation between the three variable parameters: the temperature $T$, the characteristic spacing  $\Delta\varepsilon_H =\hbar\omega_c$ between the Landau subbands, and the shift $\zeta- \varepsilon_d$ of the chemical potential $\zeta$ relative to the degeneracy energy $\varepsilon_d$. In weak magnetic fields  when   $\Delta\varepsilon_H$ is much less than the temperature $T$, the  susceptibility $\chi^{ij}$ is practically  independent of $H$. On the other hand, at $\Delta\varepsilon_H>T$ a noticeable $H$-dependence of $\chi^{ij}$ appears. In particular, at $|\zeta-\varepsilon_d|> \Delta\varepsilon_H>T$, the de Haas - van Alphen oscillations in the susceptibility occur. At higher magnetic fields, $\Delta\varepsilon_H > |\zeta- \varepsilon_d|$, $T$, i.e., in the ultra-quantum regime, these  oscillations disappear, but ${\bf M}(H)$ remains a nonlinear function of the magnetic field. Using Eqs.~(\ref{8}) and (\ref{9}), one can estimate the spacing  $\Delta\varepsilon_H\sim (e\hbar HV^2/c)^{1/2}$ and the boundary $H_T\sim cT^2/e\hbar V^2$ between the regions of the weak and strong magnetic fields where the parameter $V$ characterizes the average slope of the Dirac (Weyl) cone, $V\sim \sqrt{b_{ii}}$. Since representative values of $V$ are of the order of $10^5-10^6$ m/s \cite{wang,neupane,borisenko,jeon,z.wang,liu1}, we obtain $H_T\sim 2-200$ Oe  at $T=4$ K. In other words, at low temperatures, a noticeable dependence of $\chi^{ij}$ on $H$ can be observed for the Dirac (Weyl) semimetals even at low magnetic fields.

Below we discuss the magnetic susceptibility produced by a Dirac (Weyl) point in the weak and strong magnetic fields and in the ultra-quantum regime. Since for the Dirac and Weyl points the spectra in the magnetic field almost coincide and differ from one another by the degeneracy of the Landau subbands, the susceptibility of the Weyl point is halved as compared to the susceptibility of the Dirac point. The role of the zero Landau subbands which differ for these two cases is discussed in Sec.~\ref{sec3.3}.

\subsection{Weak magnetic fields} \label{sec3.1}

In the case of the weak magnetic fields the orbital magnetic susceptibility of the electrons described by Hamiltonian (\ref{1}), (\ref{2}) was calculated many years ago \cite{m-sv}, and a giant diamagnetic anomaly in the susceptibility was found. This anomaly is characterized by the logarithmic divergence of the susceptibility, $\chi_{ij}\propto \ln|\zeta-\varepsilon_d|$, when the chemical  potential $\zeta$ approaches $\varepsilon_d$. This anomaly was also obtained and investigated by Koshino and Ando \cite{kosh} and by Koshino and Hizbullah \cite{kosh15} for the case of the Dirac (Weyl) points without the tilt of their spectra (see also recent publications \cite{thakur,zhou18}). Finally, using the results of Ref.~\cite{m-sv,m-sh}, the magnetic susceptibility of the Dirac and Weyl points with an arbitrary tilt of their spectra was analyzed in our paper \cite{m-sh16}.

The calculation of the magnetic susceptibility $\chi^{ij}$ (per unit volume) in the region of the weak magnetic fields ($H\ll H_T$) for the case of the Dirac point leads to the following expression \cite{m-sv,m-sh16}:
\begin{eqnarray}\label{15}
 \chi^{ij}\!\!=\!-\frac{1}{6\pi^2\hbar}\!\left(\frac{e}{c}\right)^2\!\!\! \frac{\kappa^{ij}}{(b_{11}b_{22}b_{33})^{1/2}}\!\!\!\int_{0}^{\varepsilon_0} \!\!\frac{d\varepsilon}{\varepsilon}[f(-\varepsilon)\!-\!\!f(\varepsilon)],
 \end{eqnarray}
where  $\kappa^{ij}$ is given by Eq.~(\ref{9}), $f(\varepsilon)$ is the Fermi function with the chemical potential $\zeta$,
\begin{eqnarray}\label{16}
  f(\varepsilon)=\left[1+\exp\!\left(\frac{\varepsilon+\varepsilon_d -\zeta}{T}\right)
  \right ]^{-1},
 \end{eqnarray}
and $\varepsilon_0$ is a sufficiently high energy specifying the interval ($\varepsilon_d-\varepsilon_0, \varepsilon_d+\varepsilon_0$)  in  which  Hamiltonian (\ref{1})--(\ref{4}) is valid. Different choices of $\varepsilon_0$ are equivalent to a change of the   background term $\chi^{ij}_0$. Formula (\ref{16}) permits one to analyze dependences of the magnetic susceptibility on the temperature and the chemical potential that can shift by doping and deformations of the crystal.

\begin{figure}[t] 
 \centering  \vspace{+9 pt}
\includegraphics[scale=.80]{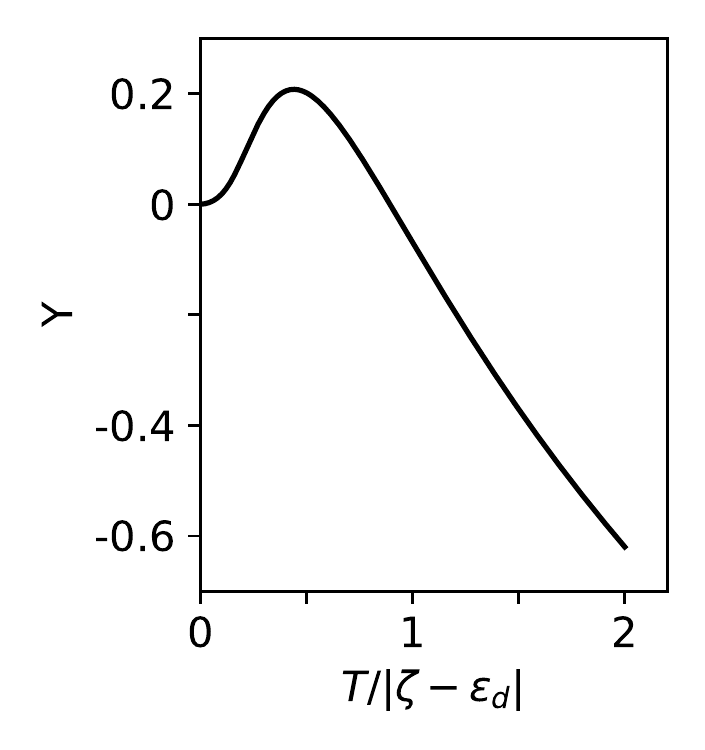}
\caption{\label{fig2} The function $Y\!(\!\frac{T}{|\zeta-\varepsilon_d}|)$ defined by Eq.~(\ref{20}).
The maximum value of $Y\approx 0.208$ is reached at $T/|\zeta-\varepsilon_d| \approx 0.443$.
 } \end{figure}

Calculating the integral in Eq.~(\ref{15}) in the limit $T\ll |\zeta- \varepsilon_d|$, we arrive at the following $\zeta$-dependence of the susceptibility:
\begin{eqnarray}\label{17}
 \chi^{ij}\!\!=\!-\frac{1}{6\pi^2\hbar}\!\left(\frac{e}{c}\right)^2\!\!\! \frac{\kappa^{ij}}{(b_{11}b_{22}b_{33})^{1/2}}\ln\left( \frac{\varepsilon_0}{|\zeta-\varepsilon_d|}\right)\!.
 \end{eqnarray}
At $|\zeta-\varepsilon_d|\lesssim T$ the divergence of $\chi^{ij}(\zeta)$ in Eq.~(\ref{17}) is truncated, and at $\zeta=\varepsilon_d$ equation (\ref{15}) yields,
\begin{eqnarray}\label{18}
 \chi^{ij}(\varepsilon_d,T)\!\approx\!-\frac{1}{6\pi^2\hbar}\!\left( \frac{e}{c}\right)^2\!\!\! \frac{\kappa^{ij}}{(b_{11}b_{22}b_{33})^{1/2}}\ln\left( \frac{\varepsilon_0}{0.88T}\right)\!.
 \end{eqnarray}
In order to analyze the temperature dependence of the susceptibility at a fixed value of $\zeta-\varepsilon_d\neq 0$, it is convenient to consider the difference $\chi^{ij}(\zeta,T)-\chi^{ij}(\zeta,0)$,
\begin{eqnarray}\label{19}
 \chi^{ij}(\zeta,T)-\chi^{ij}(\zeta,0)\!=\!-\frac{1}{6\pi^2\hbar}\!\left( \frac{e}{c}\right)^2\!\!\! \frac{\kappa^{ij}}{(b_{11}b_{22}b_{33})^{1/2}} \ Y\!\!\left(\!\frac{T}{|\zeta-\varepsilon_d|}\!\right),
 \end{eqnarray}
where the function $Y(x)$ is defined as follows:
\begin{eqnarray}\label{20}
 Y\!\!\left(\!\frac{T}{|\zeta-\varepsilon_d|}\!\right)\!=\!
\int_{0}^{\varepsilon_0} \!\!\frac{d\varepsilon}{\varepsilon}[f(-\varepsilon)\!-\!\!f(\varepsilon)] - \ln\left( \frac{\varepsilon_0}{|\zeta-\varepsilon_d|}\right).
 \end{eqnarray}
When $\varepsilon_0\gg T, |\zeta-\varepsilon_d|$, this function shown  in Fig.~\ref{fig2} is independent of the parameter  $\varepsilon_0$. It follows from Fig.~\ref{fig2} that the temperature dependence of the susceptibility is not monotonic, and its minimum  is reached at $T\approx 0.443|\zeta-\varepsilon_d|=0.443\cdot S_{\rm max}/(\pi m_*)$, see Eqs.~(\ref{25}). Interestingly, a minimum in the temperature dependence of the magnetic susceptibility of TaAs was really observed at $T\approx 185$ K \cite{liu-JMMM}. However, to describe quantitatively the experimental data, one should take into account that two types of the Weyl points with different $|\zeta-\varepsilon_d|$ exist in TaAs \cite{arnold16}.

It is also instructive to compare Eqs.~(\ref{17})--(\ref{19}) with the appropriate results for a semimetal or a doped semiconductor with a narrow gap $2\Delta_{\rm min}$ and a strong spin-orbit interaction \cite{m-sh}. As an example we refer to bismuth-antimony alloys with the magnetic field lying near the bisectrix direction \cite{buot72,bra77,m-sh00,fuseya}. In this case the divergence in Eq.~(\ref{17}) is truncated at $\zeta -\varepsilon_d\sim \Delta_{\rm min}$ where $\varepsilon_d$ now marks the middle of the gap. The $\chi^{ij}(\varepsilon_d,T)$ is practically independent of $T$ when $T< \Delta_{\rm min}$, while at  $T> \Delta_{\rm min}$ formula (\ref{18}) becomes approximately valid again. The temperature dependence of the susceptibility at $\zeta-\varepsilon_d > \Delta_{\rm min}$ still has a minimum \cite{bra77,m-sh00}, but its depth and position change with decreasing ratio $(\zeta-\varepsilon_d)/\Delta_{\rm min}$. When this ratio is close to unity, the minimum almost disappears. Thus the small gap practically does not manifest itself in the susceptibility except for the case $|\zeta-\varepsilon_d|/\Delta_{\rm min}\lesssim 1$.

It is necessary to emphasize that formulas (\ref{15})--(\ref{19}) describing the giant anomaly in the susceptibility  are valid only under the condition $\tilde a^2<1$. If $\tilde a^2> 1$, the appropriate $\chi^{ij}$ proves to be a constant \cite{m-sv}. This constant is independent of $\zeta$ and $T$ and can be incorporated into the background term. Thus the giant anomaly in the magnetic susceptibility exists only for type-I Dirac and Weyl semimetals for which the electron topological transition occurs at $\zeta= \varepsilon_d$. In the case of type-II topological semimetals the anomaly is absent. Since the volume of the electron (hole) ellipsoid near a type-I Weyl (Dirac) point is equal to
 \[
\frac{4\pi}{3}\frac{|\zeta- \varepsilon_d|^3}{(b_{11}b_{22}b_{33})^{1/2} (1-\tilde a^2)^2},
 \]
the charge carrier density $n\propto |\zeta-\varepsilon_d|^3$, and the appropriate density of states is small, $dn/d\varepsilon \propto (\zeta-\varepsilon_d)^2$. Therefore, contributions of the point to those physical quantities that are proportional to $dn/d\varepsilon$ are small, too. However, the {\it orbital} part of the magnetic susceptibility does not refer to such quantities since it is determined by virtual interband transitions of electrons under the action of the magnetic field. It is these virtual transitions between the close bands $\varepsilon_c({\bf p})$ and $\varepsilon_v({\bf p})$ that lead to the giant anomaly, and so measurements of the susceptibility can be the effective way of investigating the type-I  Dirac and Weyl points.

\begin{figure}[t] 
 \centering  \vspace{+9 pt}
\includegraphics[scale=.80]{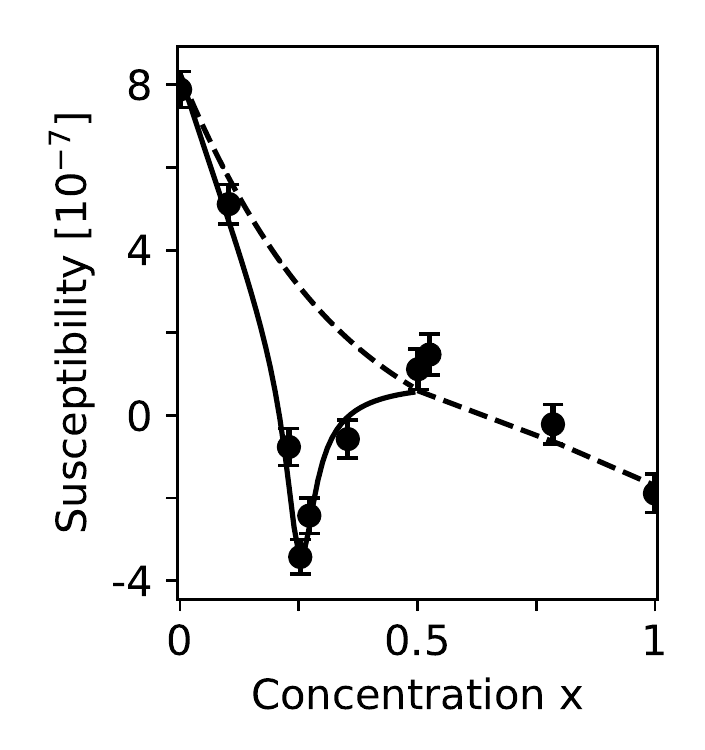}
\caption{\label{fig3} The magnetic susceptibility of Na$_{1-x}$Bi$_x$ alloys \cite{rob} (recalculated per unit volume) versus Bi concentration at the temperature $900^\circ$ C (solid circles). The dashed line depicts the background susceptibility $\chi_0$ taken here; see the text. The solid line shows $\chi=(\chi^{11}+\chi^{22}+\chi^{33})/3$ calculated with Eq.~(\ref{15}) and superimposed on $\chi_0$. The coefficient before the integral in the appropriate expression for $\chi$ is equal to $2\cdot 10^{-7}$ according to the data of Ref.~\cite{z.wang}; $\zeta(x)-\varepsilon_d=9.1(x-0.25)$ eV; $\varepsilon_0=2.6$ eV.
 } \end{figure}

The giant anomaly in the susceptibility seems to reveal itself in the experimental data obtained many years ago \cite{rob}. In that paper
a dependence of the magnetic susceptibility $\chi$ of the liquid alloys Na$_{1-x}$Bi$_x$ on the concentration $x$ of bismuth was measured at the temperature $900^\circ$ C, and a noticeable diamagnetic deep on a smooth background was observed at the concentration $x=0.25$ which corresponds to the stoichiometric formula Na$_3$Bi, Fig.~\ref{fig3}. This result can be qualitatively understood with Eq.~(\ref{15}) if one considers the $x$-dependence of the susceptibility near the concentration $x=0.25$ as the dependence of $\chi=(\chi^{11}+\chi^{22}+\chi^{33})/3$ on the chemical potential $\zeta$ for the two identical Dirac points in Na$_3$Bi. Since the electron band structure of the alloys undergoes an essential transformation with changing $x$ from $0$ to $1$, the background susceptibility $\chi_0$ cannot be considered as a constant in the whole interval of $x$. In our analysis we take $\chi_0$ as a smooth function that approximates the total susceptibility in the interval $0.5\le x \le 1$ and approaches the susceptibility of Na at $x=0$; see the dashed line in Fig.~\ref{fig3}. The function $\chi(\zeta)$ found with Eq.~(\ref{15}) is superimposed on this background; the solid line in Fig.~\ref{fig3}. In this calculation the parameters $\kappa^{ii}$ and $b_{ii}$ have been found from the data of Ref.~\cite{z.wang}, and we have used  $\zeta(x)-\varepsilon_d=9.1(x-0.25)$ eV and $\varepsilon_0=2.6$ eV as adjustable parameters.

So far we have discussed only the orbital part of the magnetic susceptibility. In the magnetic field $H$ the total electron Hamiltonian ${\mathcal H}_{\rm tot}$ is the sum of the orbital Hamiltonian (\ref{1}), (\ref{2}) in which ${\bf p}$ is replaced by ${\bf p}+e{\bf A}/c$ with ${\bf A}$ being the vector potential and the spin term ${\mathcal H}_{\rm s}=(e\hbar/mc){\bf s}\cdot{\bf H}$, i.e.,  ${\mathcal H}_{\rm tot}={\mathcal H}_{\rm or}+{\mathcal H}_{\rm s}$. In order to obtain the magnetic susceptibility, the appropriate thermodynamic potential has to be calculated in the second order in $H$. Then, this potential consists of the three parts determined by $({\mathcal H}_{\rm or})^2$, ${\mathcal H}_{\rm or}{\mathcal H}_{\rm s}$, $({\mathcal H}_{\rm s})^2$, respectively. The first part gives the orbital magnetic susceptibility $\chi_{\rm or}$ that has been analyzed above. The third and second parts specify the spin susceptibility $\chi_{\rm s}$ and the so-called  cross susceptibility $\chi_{\rm s-o}$, respectively. For the Dirac and Weyl points the $\chi_{\rm s}$ and $\chi_{\rm s-o}$ were calculated in Ref.~\cite{kosh15}, see also Refs.~\cite{ominato,ominato1}. It is important that $\chi_{\rm s}$ and $\chi_{\rm s-o}$ are relatively small when $\zeta$ lies near $\varepsilon_d$ \cite{kosh15}:
 \[\
 |\chi_{\rm s-o}|\sim \eta|\chi_{\rm or}|,\ \ \  |\chi_{\rm s}| \sim \eta^2 |\chi_{\rm or}|,
 \]
where the small dimensionless parameter $\eta$ is the ratio of the spin magnetic moment $\mu_B$ to the characteristic orbital magnetic moment $(g/2)\mu_B$ of an electron in the bands ``$c$'' and ``$v$'', $\mu_B=e\hbar/2mc$ is the Bohr magneton,  and $g$ is the appropriate electron $g$ factor. Taking into account the results of  Sec.~\ref{sec2.3}, one has $g=2m/m_*$ and
\begin{eqnarray} \label{21}
\eta\sim \frac{m_*}{m}\sim \frac{(\zeta-\varepsilon_d)}{mV^2},
 \end{eqnarray}
where $m_*$ is the cyclotron mass of the charge carriers in the bands ``$c$'' and ``$v$'', and $V$ is the average slope of the Dirac (Weyl) cone. The smallness of $\chi_{\rm s}$ and $\chi_{\rm s-o}$ permits  one to disregard these parts of the susceptibility in the vicinity of the Dirac and Weyl points.

\subsection{Strong magnetic fields}\label{sec3.2}

In the case of the strong magnetic fields, $H\gg H_T$, the  magnetization ${\bf M}$ of the electrons described by Hamiltonian (\ref{1}), (\ref{2}) was calculated in Ref.~\cite{m-sh}, and the   magnetic-field dependence $M\propto H\ln H$ was found for the ultra-quantum regime when   $\Delta\varepsilon_H> |\zeta-\varepsilon_d|$, $T$. By analogy with Eq.~(\ref{18}), this dependence can be qualitatively understood if one truncates the divergence in formula (\ref{17}) at $|\zeta-\varepsilon_d|\sim \Delta \varepsilon_H$ (there is also a parallel between the contribution $\propto H^2\ln H$ to the $\Omega$ potential and the appropriate  result of the $3+1$ quantum electrodynamics with massless fermions \cite{niss18}). This logarithmic in $H$ factor in the magnetization was also obtained in the papers \cite{roy,moll16} for the case of the isotropic Dirac (Weyl) points. Finally, using the results of Ref.~\cite{m-sh}, the magnetization of the Dirac and Weyl points with an arbitrary tilt of their spectra was analyzed in our recent paper \cite{m-sh16}. For the magnetic-field region $|\zeta-\varepsilon_d|>\Delta\varepsilon_H>T$, the de Haas - van Alphen oscillations associated with these points were theoretically investigated in Refs.~\cite{m-sh,wright,ashby,wang-prl16,m-sh16}.

\begin{figure}[tbp] 
 \centering  \vspace{+9 pt}
\includegraphics[scale=.99]{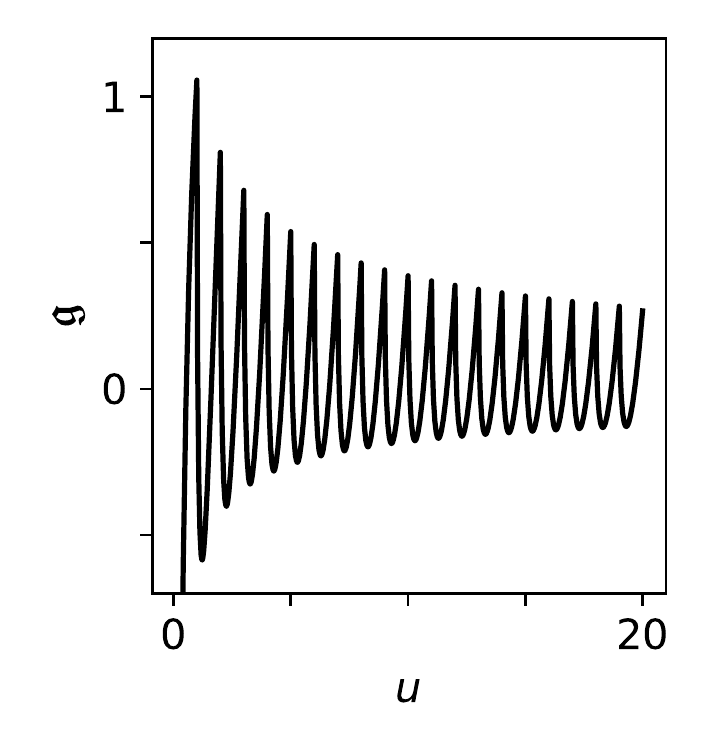}
\caption{\label{fig4} The function $\mathfrak{g}(u)$, Eq.~(\ref{24}), that describes the oscillations of the magnetization \cite{m-sh16}.
At integers $u$ the function $\mathfrak{g}(u)$ exhibits sharp peaks.
 } \end{figure}   

The calculation of the magnetization in the region of the strong magnetic fields ($H\gg H_T$) for the case of the Dirac points gives \cite{m-sh,m-sh16}:
\begin{eqnarray}\label{22}
M_i\!=\!H\!\sum_{j=1}^{3}\!\chi^{ij}_{H\to 0}\!\!\cdot\!n_j\!+\!\frac{e Q_i(\zeta- \varepsilon_d)^2  \mathfrak{g}(u)}{12\pi^2\!\hbar^2 c(b_{11}b_{22}b_{33})^{\!1\!/2}\!(1-\tilde a^2)R_n^{1/2}},~~
 \end{eqnarray}
where $\chi^{ij}_{H\to 0}$ is the susceptibility in the weak magnetic fields, Eq.~(\ref{17}); ${\bf n}$ is the unit vector along the magnetic field, $H_j=n_jH$; $Q_i=\sum_j\kappa^{ij} n_j$; $\kappa^{ij}$ and  $R_n$ are given by Eqs.~(\ref{9}), and
\begin{equation}\label{23}
 u\equiv  \frac{(\zeta-\varepsilon_d)^2 c}{2e\hbar(1-\tilde a^2)R_n^{1/2}H}=\frac{S_{\rm max}c}{2\pi\hbar e H}.
 \end{equation}
Here $S_{\rm max}$ is the area of the maximal cross-section of the constant-energy surface $\varepsilon_{c,v}({\bf p})=\zeta$ by the plane perpendicular to the magnetic field ($p_n=$ const.); the maximum is found relative to $p_n$ that is the quasi-momentum along the magnetic field. The universal function $\mathfrak{g}(u)$ is independent of the parameters of the Dirac point:
\begin{eqnarray}\label{24}
\mathfrak{g}(u)=-\frac{\ln(2\sqrt u) +A-\frac{1}{4}}{u}+3+
  \frac{6}{u}\sum_{m=1}^N\Bigg[\sqrt{u(u-m)} \nonumber \\
-2m\ln\left(\frac{\sqrt{u}+\sqrt{u-m}}{\sqrt{m}} \right)\Bigg],~~~~
 \end{eqnarray}
where $A\approx 1.50$, and $N\equiv [u]$ is the integer part of $u$.  This $N$ is the number of the Landau subbands occupied by electrons in the conduction band or by holes in the valence band. The function $\mathfrak{g}(u)$ for not-too-high $u$ is shown in Fig.~\ref{fig4}, while at $u\gg 1$ it is given by the expression:
\begin{eqnarray}\label{25}
\mathfrak{g}(u)\approx -\frac{6}{u^{1/2}}\zeta(-1/2,\{u\})=-\frac{3}{\pi\sqrt{2u}} \sum_{n=1}^{\infty}\frac{1}{n^{3/2}}\,\sin\!\!\left(2\pi nu-\frac{\pi}{4}\right),
 \end{eqnarray}
where $\{u\}=u-[u]$, and $\zeta(s,a)$ is the Hurwitz zeta function.
Formula (\ref{22}) enables one to analyze both the de Haas - van Alphen oscillations of the magnetization and the $H$-dependence of ${\bf M}$ in the ultra-quantum regime.

Using $M_i(\zeta-\varepsilon_d,H,T=0)$ given by Eq.~(\ref{22}), one can find the magnetization $M_i(\zeta-\varepsilon_d,H,T)$ at an arbitrary temperature,
\begin{equation}\label{26}
M_i(\zeta-\varepsilon_d,H,T)=-\int_{\!-\varepsilon_0}^{\varepsilon_0}\!\! d\!\varepsilon\ M_{i}(\varepsilon,H,0) f'\!(\varepsilon),
 \end{equation}
where $f'(\varepsilon)$ is the derivative of the Fermi function defined by formula (\ref{16}). It follows from this expression that the second (oscillating) term in Eq.~(\ref{22}) is suppressed at temperatures when $H_T \gg H$, and one again arrives at formula (\ref{15}).

\subsubsection{Oscillations}

Consider now the longitudinal magnetization $M_{\parallel}=\sum_{i} M_i n_i$ in more detail. Taking into account the expressions for the maximal cross-section area $S_{\rm max}$ of the Fermi surface, the cyclotron mass $m_*$, and $|S''|$ corresponding to this cross section,
\begin{eqnarray}\label{27}
S_{\rm max}=\frac{\pi(\zeta-\varepsilon_d)^2}{R_n^{1/2}(1-\tilde a^2)}, \ \ m_*\equiv \frac{1}{2\pi}\frac{\partial S}{\partial \varepsilon}=\frac{(\zeta-\varepsilon_d)}{R_n^{1/2}(1-\tilde a^2)}, \nonumber \\
|S''|\equiv |\frac{\partial^2 S}{\partial p_n^2}|=\frac{2\pi b_{11}b_{22}b_{33}(1-\tilde a^2)}{R_n^{3/2}},\ \ \sum_{i}Q_in_i=R_n,
 \end{eqnarray}
we obtain the following formula for $M_{\parallel}$ from Eq.~(\ref{22}):
\begin{eqnarray}\label{28}
M_{\parallel}\!=\!\chi_{\parallel}(H\!\!\to\! 0)\!\cdot\! H\!\!+\!C\cdot\mathfrak{g}(u),
 \end{eqnarray}
where $\chi_{\parallel}(H\!\!\to\! 0)=\sum_{i,j} \chi^{ij}_{H\to 0}n_in_j$ is the longitudinal magnetic susceptibility in weak magnetic fields, and the coefficient $C$ before the oscillating function $\mathfrak{g}(u)$ looks like
\begin{eqnarray}\label{29}
C&=&\frac{e S_{\rm max}^{3/2}}{6\sqrt 2\pi^3\!\hbar^2 c|m_*||S''|^{1/2}}.
 \end{eqnarray}
For $u\gg 1$, i.e., for $\zeta-\varepsilon_d \gg \Delta \varepsilon_H$,  the second term in formula (\ref{28}) describes the well-known de Haas - van Alphen oscillations at $T=0$, with the familiar Lifshitz-Kosevich formula \cite{Sh} following from Eq.~(\ref{25}). Note that in the case of the Dirac points the so-called spin factors of this formula are now expressed via the phases of the harmonics (see below). The thermal damping factors of the formula are obtained with Eq.~(\ref{26}).

In general the de Haas - van Alphen oscillations associated with an arbitrary extremal cross-section of a metal have the following functional form: $u^{-1/2}G(u-\phi)$ \cite{shen} where $G(x)$ is a periodic function, and the phase of the oscillations $\phi$ is determined by the subleading-order term in the quantization condition, i.e., by  $\gamma$ or the $g$ factor. Specifically,  $\phi=\gamma$ for the nondegenerate band, cf. Eq.~(\ref{12}), whereas    Eq.~(\ref{11}) gives $\phi=1/2\pm (gm_*/4m)$ in the case of the doubly degenerate  electron band (in the Lifshitz-Kosevich formula the part of the phase $\pm gm_*/4m$ is usually written as the spin factor). Comparing relation (\ref{7}) with the general quantization condition given by Eq.~(\ref{11}) or Eq.~(\ref{12}), one finds $\phi=0$ for the Weyl and Dirac semimetals, and this conclusion also follows directly from formula (\ref{28}). As a result, in the topological semimetals the phase of the oscillations is shifted as compared to the conventional metals. This phase shift is the characteristic property of the Dirac (Weyl) fermions. However, it is necessary to keep in mind that the same phase shift should be observed in a Dirac semimetal with a small induced gap  (i.e.,  in a doped narrow-gap semiconductor with a strong spin-orbit interaction) since in this case the spectrum in the magnetic field is still described by Eq.~(\ref{7}), see the end of Sec.~\ref{sec2.2}.

At low temperatures, $T\ll \Delta \varepsilon_H$, the phase of the oscillations in $M_{\parallel}$ can be experimentally determined with the so-called Landau-level fan diagram \cite{Sh}, i.e., with plotting the dependence $1/H_l$ versus $l$ where $H_l$ is the magnetic field corresponding to $l$-th peak in the magnetization. This peak occurs when the $l$-th Landau-subband edge crosses the Fermi level, Fig.~\ref{fig4}. However, with increasing temperature  the sharp peaks in the magnetization are smoothed, and these distinct markers of the subband edges die out. Besides, in real semimetals several electron orbits can contribute to the oscillations. In this case the phase of the oscillations is usually found with the Fourier analysis of the experimental data. The $n$-th harmonic $\Delta M_n$ of $M_{\parallel}$ looks like \cite{Sh}:
\begin{eqnarray} \label{30}
 \Delta M_n \propto \frac{H^{1/2}}{n^{3/2}}\sin\!\left(\!\!2\pi n \!\left[\frac{F}{H}-\phi\!\right] \pm \frac{\pi}{4}\right),
 \end{eqnarray}
where $F=S_{\rm ex}c/2\pi\hbar e$ is the frequency determined by the extremal cross-section area $S_{\rm ex}$ of the Fermi surface, and the additional offsets  $\pm \pi/4$ refer to the minimal and maximal $S_{\rm ex}$, respectively. This additional offset is due to the expansion of the function $G(x)$ in the Fourier series, cf. Eq.~(\ref{25}). Approximating the appropriate experimental data by a number of these harmonics, one can find their phase, i.e.,  $\gamma$ or the $g$ factor. The same considerations are applicable to the Shubnikov - de Haas oscillations of the conductivity $\sigma$ measured in the plane perpendicular to the magnetic field. Its $n$-th harmonic $\Delta \sigma_n$ is described by the formula that is similar to Eq.~(\ref{30}) \cite{LL10}:
\begin{eqnarray} \label{31}
 \Delta \sigma_n \propto \frac{1}{n^{1/2}H^{3/2}}\cos\!\left(\!\!2\pi n\!\left[\frac{F}{H}-\phi \right] \pm \frac{\pi}{4}\right).
 \end{eqnarray}
Note that the use of the resistivity rather than conductivity in analyzing the oscillations can lead to an inaccurate determination of $\phi$ \cite{ando13}. At present the phase of the oscillations has been experimentally investigated for a number of the Weyl and Dirac semimetals \cite{huang15,luo15,hu16,du16,sergelius,wang16,he-r(h),pari,zhao15,nara15,xiang15,desr,cao15,he16}.
and values of $\phi$ lying between $0$ and $1/2$ were obtained in papers \cite{du16,wang16,pari,zhao15,cao15}.

Let us briefly outline possible reasons of the deviation of $\phi$ from $0$ or $1/2$ in the experiments. First, when $H$ approaches the boundary $H_1$ of the ultra-quantum regime, see Sec.~\ref{sec3.2.2}, the chemical potential begins noticeably depends on $H$ if there is no large electron group which stabilizes $\zeta$ in the crystal. As a result, the shape and the phase of the last oscillations in $1/H$ changes, and $\phi$ found with these oscillations may differ from the true value. Second, the magnetic breakdown \cite{Sh,azbel,alex1,alex2} can lead to any value of $\phi$ lying in the interval from $0$ to $1$. However in this case a sharp dependence of $\phi$ on the direction of the magnetic field is expected. Interestingly, angular dependences of $\phi$ were really found in Refs.~\cite{xiang15,cao15}. Third, so far we have considered the situation when the chemical potential $\zeta$ is so close to the degeneracy energy $\varepsilon_d$ that $\eta \ll 1$ where the parameter $\eta$ is defined by Eq.~(\ref{21}). If this parameter is not-too-small (but the Fermi surfaces of different Weyl or Dirac points do not merge into a single surface), an incorporation of nonlinear in ${\bf p}$ terms in Hamiltonian (\ref{1}), (\ref{2}) may become necessary. These terms  modify the spectrum (\ref{3}), (\ref{5}). Besides, in this approximation the $g$ factor is no longer equal to the universal value $2m/m_*$, i.e., $g_{c,v}=(2m/m_*)+\Delta g_{c,v}$. The corrections $\Delta g_{c,v}$ are caused not only by the spin term ${\mathcal H}_{\rm s}=(e\hbar/mc){\bf s}\cdot{\bf H}$  but also by contributions of the bands different from `$c$'' and ``$v$'' to the orbital parts of $g_{c}$ and $g_{v}$. In principle these $\Delta g_{c,v}$ can be found with a perturbation theory \cite{g3}. As to the Weyl semimetals, it was demonstrated by Wright and McKenzie \cite{wright} that the parameter $\gamma$ may differ from its value $\gamma=0$  when nonlinear in ${\bf p}$ terms appear in the Hamiltonian, and if there is a gap in the spectrum. However, because of the tilt of the spectrum, the maximal cross-section in the Weyl semimetals generally does not pass through the point ${\bf p}=0$, see Sec.~\ref{sec2.1} and Fig.~\ref{fig1}. Therefore the gap always exists for the electrons in this cross-section. In other words, one may expect to detect a nonzero $\gamma$  when the nonlinear terms in the Hamiltonian become essential, and an estimate gives $\gamma \sim \eta$.

In the case of the Dirac semimetals, if the corrections to the spectrum (\ref{3}), (\ref{5}) are known, the following generalization of Eq.~(\ref{28}) can be used to describe the oscillations in the magnetization $M_{\parallel}$ and to find the appropriate $\Delta g$ experimentally:
\begin{eqnarray}\label{32}
M_{\parallel}\!=\!\chi_{\parallel}(H\!\!\to\! 0)\!\cdot\! H\!\!+\!C\cdot\frac{\mathfrak{g}(u_+)+ \mathfrak{g}(u_-)}{2} ,
 \end{eqnarray}
where
\begin{eqnarray*}
u_{\pm}\!&=&u \pm \frac{m_* \Delta g}{4m},
 \end{eqnarray*}
and the quantities $S_{\rm ex}$, $m_*$, $S''$ defining the coefficient $C$ are calculated with the corrected spectrum. Formula (\ref{32}) shows that the nonzero $\Delta g=g-(2m/m_*)$ leads to the splitting of the peaks in the magnetization. A similar splitting of the oscillations in the resistivity was really observed in large magnetic fields \cite{zhao15,nara15,cao15} when the condition $\Delta \varepsilon_H\gg T$ is well fulfilled, and the shape of the oscillations differs from a sinusoid.

\subsubsection{The ultra-quantum regime}\label{sec3.2.2}

At $\Delta\varepsilon_H> |\zeta-\varepsilon_d|$, i.e. $u<1$, the de Haas - van Alphen oscillations in magnetization disappear, and the ultra-quantum regime occurs. From Eq.~(\ref{23}) and the condition $u=1$ we find the boundary $H_1$ of this regime,
\begin{equation}\label{33}
   H_1 =\frac{S_{\rm max}c}{2\pi\hbar e}= \frac{(\zeta-\varepsilon_d)^2 c}{2e\hbar(1-\tilde a^2)R_n^{1/2}}.
 \end{equation}
At $H>H_1$ the sum in Eq.~(\ref{24}) vanishes, and formula (\ref{22}) for the magnetization of the Dirac point reduces to the expression:
\begin{eqnarray}\label{34}
M_i\!&=&\!-\frac{1}{6\pi^2\hbar}\!\left(\frac{e}{c}\right)^2\!\!\! \frac{Q_i H}{(b_{11}b_{22}b_{33})^{1/2}}\left [\frac{1}{2}\ln\left( \frac{2\varepsilon_0^2c}{e\hbar(1-\tilde a^2)R_n^{1/2}H}\right)\!+A-\frac{1}{4}\right] \nonumber \\
&+&\frac{e Q_i(\zeta- \varepsilon_d)^2}{4\pi^2\!\hbar^2 c(b_{11}b_{22}b_{33})^{\!1\!/2}\!(1-\tilde a^2)R_n^{1/2}},
 \end{eqnarray}
We see that the magnetization in the ultra-quantum regime comprises the constant, the linear in $H$ term, and the nonlinear term which is proportional to $H\ln H$. Interestingly, the difference between this magnetization and that extrapolated from the region of the weak magnetic fields is independent of the cut-off parameter $\varepsilon_0$, and it is expressed in terms of the parameters of the Dirac point only,
 \begin{eqnarray}\label{35}
M_i\!-\!H\!\sum_{j=1}^{3}\!\chi^{ij}_{H\to 0}n_j\!=-\frac{1}{6\pi^2\hbar}\!\left(\frac{e}{c}\right)^2\!\!\!\!\! \frac{Q_i H}{(b_{11}b_{22}b_{33})^{1/2}}\!\left [\!\frac{1}{2}\ln\!\!\left(\!\frac{4H_1}{H}\!\right)\!\!+\!A-\!\frac{1}{4} -\!\frac{3H_1}{H}\right].
 \end{eqnarray}
An experimental investigation of this difference also permits one to eliminate the background susceptibility $\chi_0^{ij}$ from consideration. Using formula (\ref{34}), one can easily calculate the magnetic torque ${\bf K}$ per unit volume: ${\bf K}=[{\bf M}\times {\bf H}]$.

Discuss now the dependence of the chemical potential $\zeta$ on the magnetic field. In the regions of the weak magnetic fields and of the de Haas - van Alphen oscillations, this dependence is negligible as long as $H\ll H_1$. However, it  becomes essential at $H\sim H_1$ and in the ultra-quantum regime when $H>H_1$. In particular, equating the charge carrier densities at $H>H_1$ and at $H=0$, we arrive at
 \[
\zeta -\varepsilon_d = (\zeta_0 -\varepsilon_d)\frac{2H_1}{3H}\sim \frac{(\zeta_0 -\varepsilon_d)^3}{(\Delta \varepsilon_H)^2},
 \]
where $\zeta_0$ is the chemical potential at $H=0$, the field $H_1$ is given by formula (\ref{33}), and  $\Delta \varepsilon_H=\sqrt{e\hbar \alpha H/c}$ according to Eq.~(\ref{8}). Thus, $\zeta$ approaches the degeneracy energy $\varepsilon_d$ with increasing $H$. This $H$-dependence of $\zeta$  will suppress the last term in Eq.~(\ref{34}) which is proportional to $(\zeta_0 -\varepsilon_d)^2$. Of course, to find quantitatively this dependence for a real sample, it is necessary to take into account that different types of the Weyl (Dirac) points with different $|\zeta- \varepsilon_d|$ may exist in a topological semimetal along with Fermi pockets of the conventional charge carriers.

In recent paper \cite{moll16} Moll with coauthors discovered a pronounced anomaly in the magnetic torque of the Weyl semimetal NbAs upon entering the ultra-quantum regime. The torque changed its sign in this regime, signalling a reversal of the magnetic anisotropy. A similar change in the torque, but without the reversal of its sign,  was also  observed in TaAs \cite{zhang-nc19}. If using Eqs.~(\ref{17}) and (\ref{34}), one calculates the difference between the torque in the ultra-quantum regime and that extrapolated from the region of the weak magnetic fields, a formula closely resembling Eq.~(\ref{35}) is obtained  (only the coefficients before the square brackets are different in these expressions). This means that the torque does change in the ultra-quantum regime. However to describe the experimental data quantitatively, one must take into account several charge-carrier groups existing in these semimetals.

In order to estimate the effect of the small gap $2\Delta_{\rm min}$ in the Dirac spectrum on the magnetization in the ultra-quantum regime, it is convenient to introduce the characteristic field,
\begin{equation*}
   H_{\Delta} = \frac{\Delta_{\rm min}^2 c}{2e\hbar(1-\tilde a^2)R_n^{1/2}},
 \end{equation*}
that is similar to $H_1$ defined by Eq.~(\ref{33}). The meaning of $H_{\Delta}$ is that at $H> H_{\Delta}$ the spacing between the Landau subbands  $\Delta\varepsilon_H$ exceeds  $\Delta_{\rm min}$. If $H_1 > H_{\Delta}$, the gap has no effect on the magnetization in the ultra-quantum regime at $H>H_1$, and this magnetization  is still described by Eqs.~(\ref{34}), (\ref{35}) \cite{m-sh}. However, if $H_1 < H_{\Delta}$, the de Haas - van Alphen oscillations disappear at $H>H_1$, but the magnetization reaches its asymptotic behavior given by Eqs.~(\ref{34}), (\ref{35}) only at $H> H_{\Delta}$. In other words, one may expect that the dependence  $M_i(H)$ deviates from Eqs.~(\ref{34}), (\ref{35}) in the interval $H_1<H< H_{\Delta}$. This dependence can be found with formulas of Ref.~\cite{m-sh}. Besides, as in the case of the weak magnetic fields, the bismuth-antimony alloys with the magnetic field lying near the bisectrix direction  provide the useful example for the investigation of the effect of the gap and nonlinear in ${\bf p}$ terms in Hamiltonian on the magnetization in the ultra-quantum regime \cite{bra77,m-sh00}.

\subsection{Magnetization of the Weyl semimetals}\label{sec3.3}

In Secs.~\ref{sec3.1} and \ref{sec3.2} we have presented formulas mainly for the case of the Dirac points assuming that
the susceptibility of a Weyl node is halved as compared to the susceptibility of the appropriate Dirac one. However the zeroth Landau subbands are distinct from each other for these two types of the points, cf. Eqs.~(\ref{8}) and (\ref{10}). Let us represent a contribution of the zeroth Landau subband of a Weyl node to the $\Omega$ potential as a half of the contribution of the zeroth subband of the Dirac point plus some additional term $\delta \Omega$. At $\tilde a^2 <1$ this term has the form:
 \begin{eqnarray}\label{36}
\delta \Omega=-q_{ch}\frac{eH}{(2\pi\hbar)^2c}\! \frac{(\tilde {\bf a}\cdot\tilde{\bf n})}{(1-\tilde a^2)}\!\left (\!\frac{(\varepsilon_0+\zeta-\varepsilon_d)^2}{2}+\!\frac{\pi^2T^2}{6}\!\right),
 \end{eqnarray}
while at $\tilde a^2 >1$ it looks like
 \begin{eqnarray}\label{37}
\delta \Omega=-q_{ch}\frac{eH{\rm sign}(\tilde {\bf a}\cdot \tilde{\bf n})}{(2\pi\hbar)^2c}\! \frac{\sqrt{(1-\tilde a^2)\tilde{\bf n}^2+(\tilde {\bf a}\cdot \tilde{\bf n})^2}}{(\tilde a^2-1)}\!\left (\!\frac{(\varepsilon_0+\zeta - \varepsilon_d)^2}{2} +\!\frac{\pi^2T^2}{6}\!\right),
 \end{eqnarray}
where we have used the notations of Eqs.~(\ref{9}).
This $\delta \Omega$ is the linear function of the magnetic field, and therefore it leads to the term ${\bf M}_{\rm sp}=-\partial(\delta\Omega)/\partial {\bf H}$ which is independent of $H$, i.e., to a spontaneous magnetization.

However in Weyl semimetals, the Weyl nodes always occur in pairs of opposite chirality $q_{ch}$, and they cannot exist if both the time reversal and inversion symmetries are present in a crystal  \cite{armit}. Let a Weyl node of the chirality $q_{ch}$ and with the tilt of the spectrum defined by the vector ${\bf a}$ be at the point ${\bf p}_0$ of the Brillouin zone. Consider the case when a crystal possesses the time reversal symmetry while the inversion symmetry is broken. The semimetals of the TaAs-family and the type-II semimetal WTe$_2$ just fall into this class. In this case another Weyl node of the same chirality and with the opposite ${\bf a}$ exists at the point $-{\bf p}_0$, see e.g. the supplemental material of Ref.~\cite{yu16}. Thus the terms $\delta \Omega$ of these two Weyl points always cancel each other, and one may disregard them. In this case the magnetization of a Weyl point really can be calculated as a half of the magnetization for the appropriate Dirac point. Another situation takes place if there is a center of the inversion in a crystal, but the time reversal symmetry is broken. This situation occurs in magnetic Weyl semimetals. In this case the Weyl point at $-{\bf p}_0$ has the opposite vector ${\bf a}$ and chirality. Hence $\delta \Omega({\bf p}_0)=\delta \Omega(-{\bf p}_0)$, and the spontaneous magnetization generated by the Weyl points does may appear. Then it will renormalizes the magnetization of the semimetal, and the parameters of the Weyl points are established self-consistently. To gain insight into the magnitude of the spontaneous magnetization, $M_{\rm sp}$, generated by a Weyl point, let us take $\varepsilon_0 = 1$ eV (assuming that $T$, $\zeta- \varepsilon_d \ll \varepsilon_0$), $\tilde a^2=0.5$, and $b_{11}=b_{22}=b_{33}=V^2$ with $V= 10^6$ m/s. Then we obtain the following estimate from Eq.~(\ref{36}): $M_{\rm sp}\approx \tilde a(1-\tilde a^2)^{-1}\varepsilon_0^2 m \mu_B/(4\pi^2 \hbar^3 V) \approx 7\cdot 10^{20}\mu_B/{\rm cm}^3$ where $\mu_B$ is the Bohr magneton.

\section{Nodal-line semimetals}\label{line}

\begin{figure}[tbp] 
 \centering  \vspace{+9 pt}
\includegraphics[scale=.95]{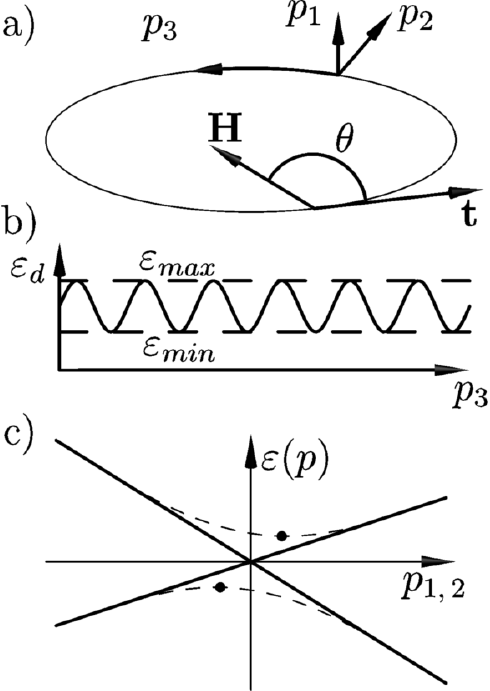}
\caption{\label{fig5} (a) The nodal line that has the shape of a ring; ${\bf t}$ is the unit tangent vector to the line at one of its points; $\theta$ is the angle between this ${\bf t}$ and the magnetic field ${\bf H}$; the coordinate $p_1$-$p_2$-$p_3$ are shown near a point of the line. (b) A dependence of the degeneracy energy $\varepsilon_d$ on the coordinate $p_3$ along the line; $\varepsilon_{min}$ and $\varepsilon_{max}$ are the minimum and maximum values of $\varepsilon_d(p_3)$. (c) The electron dispersion relation (the solid lines) in the vicinity of the nodal line in the $p_1$-$p_2$ plane perpendicular to this line. The dashed lines show this relation when the spin-orbit interaction leads to the gap $2\Delta_0$ at the point $p_1=p_2=0$. In this case the minimal indirect gap $2\Delta_{\rm min}=2\Delta_0 (1-\tilde a_{\perp}^2)^{1/2}$ exists in the spectrum. The black dots mark the minimum and maximum of the conduction and valence bands, respectively.
 } \end{figure}

As was mentioned in the Introduction, the band-contact lines in crystals always have a nonzero difference $\varepsilon_{max} -\varepsilon_{min}\equiv 2\Delta$ between the maximum and minimum band-degeneracy energies. However this difference in the nodal-line semimetals is relatively small as compared to the inherent energy scale of  crystals ($1-10$ eV). Such nodal lines  exist in  rhombohedral graphite  \cite{hei1,pie,mcclure,kopnin},  three-dimensional graphene networks \cite{weng}, Ca$_3$P$_2$ \cite{xie},  Cu$_3$NPd \cite{kim,yu}, CaAgP \cite{yama}, ZrSiS  \cite{schoop,neupane1}, ZrSiTe \cite{schoop1}, alkaline-earth germanides and silicides \cite{huang1}, PbTaSe$_2$ \cite{bian}, and SrIrO$_3$ \cite{chen}. In the most of these nodal-line semimetals the spin-orbit interaction is weak, and the band-contact lines generally exist only if one neglects this interaction. The spin-orbit coupling lifts the degeneracy of the conduction and valence bands along the line, and a small gap between the bands appears, Fig.~\ref{fig5}. The effect of this gap on the magnetic susceptibility was studied in Refs.~\cite{m-sv} and \cite{m-sh}, and it was found that this effect, as a rule, is negligible. Because of this, we mainly neglect this gap below.

\subsection{Electron spectrum near a band-contact line} \label{sec4.1}

In the vicinity of a band-contact line along which the conduction and valance bands touch, let us introduce orthogonal curvilinear coordinates so that the axis ``$3$'' coincides with the line. The axes ``$1$'' and ``$2$'' are perpendicular to the third axis at every point of the band-contact line, and the appropriate coordinate $p_1$ and $p_2$ are measured from this line, Fig.~\ref{fig5}. In these coordinates, the electron spectrum near the line has the form  \cite{m-sh,m-sh14}:
\begin{eqnarray}\label{38}
 \varepsilon_{c,v}\!\!&=&\!\varepsilon_d(p_3)\!+\!{\bf a}_{\perp}\cdot {\bf p}_{\perp}\pm E_{c,v},\\
 E_{c,v}^2\!\!&=&\!\Delta_0^2+b_{11}p_1^2+b_{22}p_2^2, \nonumber
 \end{eqnarray}
where $\varepsilon_d(p_3)$ describes a dependence of the  degeneracy energy along the line (the $\varepsilon_{max}$ and $\varepsilon_{min}$ mentioned above are the maximum and minimum values of the function  $\varepsilon_d(p_3)$), Fig.~\ref{fig5}; ${\bf p}_{\perp}=(p_1,p_2,0)$ and ${\bf a}_{\perp}=(a_1,a_2,0)$ are the vectors perpendicular to the line at every point of it; the parameters of the spectrum  $b_{11}$, $b_{22}$, and ${\bf a}_{\perp}$ generally depend on $p_3$. It is implied here that the directions of the axes ``1'' and ``2'' are chosen so that the quadratic form $E_{c,v}^2$ is diagonal (these directions generally changes along the line). For completeness, in Eq.~(\ref{38}) we also take into account the gap $\Delta_0=\Delta_0(p_3)$ induced by the spin-orbit interaction along the line, Fig.~\ref{fig5}. However this gap, as a rule, is implied to be very small, $\Delta_0 \to 0$. The vector ${\bf a}_{\perp}(p_3)$ specifies the tilt of the Dirac spectrum in the $p_1$-$p_2$ plane which is perpendicular to the line at the point $p_3$. Below we shall consider only the case when the length of the vector $\tilde{\bf a}_{\perp}\equiv (a_1/\sqrt{b_{11}}, a_2/\sqrt{b_{22}},0)$ is less than unity,
\begin{eqnarray*}
 \tilde a_{\perp}^2=\frac{a_1^2}{b_{11}}+\frac{a_2^2}{b_{22}}<1,
 \end{eqnarray*}
since at $\tilde a_{\perp}^2>1$ the magnetic susceptibility does not exhibit any essential anomaly in its dependences on $\zeta$, $H$, and $T$ \cite{m-sv,m-sh,m-sh16}. This means that if $\tilde a_{\perp}^2>1$ for a part of the band-contact line, this part should be disregarded in all subsequent formulas.

\begin{figure}[tbp] 
 \centering  \vspace{+9 pt}
\includegraphics[scale=.95]{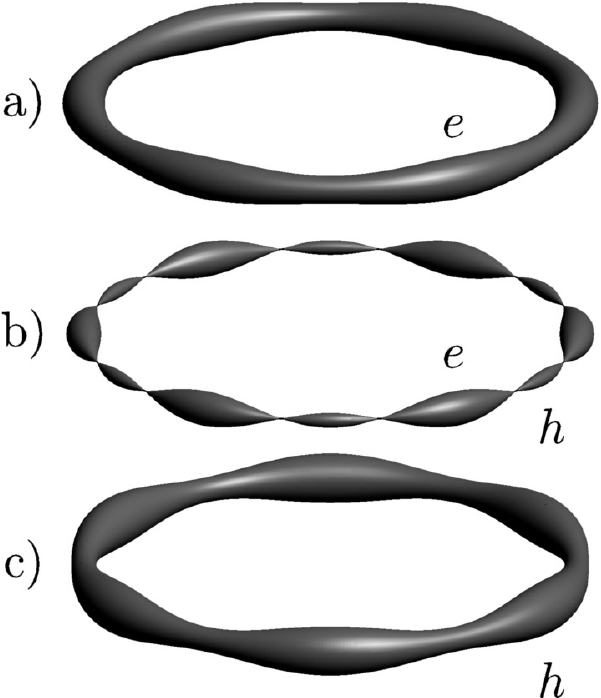}
\caption{\label{fig6} The Fermi surface in the nodal-line semimetal with the band-contact line shown in Fig.~\ref{fig5} at $\zeta>\varepsilon_d^0+\Delta$ (a), $\varepsilon_d^0+ \Delta>\zeta>\varepsilon_d^0-\Delta$ (b), and $\varepsilon_d^0- \Delta>\zeta$ (c). Letters e and h indicate the electron and hole types of the Fermi surface, respectively.
 } \end{figure}   

When the parameter $\Delta \equiv (\varepsilon_{max} -\varepsilon_{min})/2$ is small as compared to the characteristic scale $\varepsilon_0$ of electron band structure (i.e., $\Delta\ll \varepsilon_0\sim 1-10$ eV) and $\tilde a_{\perp}^2<1$, the Fermi surface $\varepsilon_{c,v}({\bf p}_{\perp},p_3)=\zeta$ of the semimetal looks like a narrow corrugated electron or hole tube for $\zeta-\varepsilon_d^0\gtrsim \Delta$ or $\zeta- \varepsilon_d^0 \lesssim -\Delta$, respectively, Fig.~\ref{fig6}.  Here $\varepsilon_d^0\equiv (\varepsilon_{max}+ \varepsilon_{min})/2$ is the mean degeneracy energy of the nodal line. The band-contact line lies inside this tube. If  $|\zeta- \varepsilon_d^0| <  \Delta$, the Fermi surface has a self-intersecting shape and consists of the electron and hole pockets touching at some points of the line, i.e., it looks like ``link sausages'', Fig.~\ref{fig6}. Thus, if the chemical potential $\zeta$ decreases and passes through the critical energies $\varepsilon_{max}=\varepsilon_d^0 + \Delta$ and $\varepsilon_{min}=\varepsilon_d^0 - \Delta$, the following two electron topological transitions of $3\frac{1}{2}$ order occur \cite{m-sh14,m-sh-jltp}: The electron tube first transforms into the self-intersecting Fermi surface and then this surface transforms into the hole tube. We shall assume below that all transverse dimensions of the Fermi-surface tubes and pockets, which are of the order of  $|\zeta- \varepsilon_d(p_3)|/V$  where $V\sim ({b_{11}b_{22}})^{1/4}$, are small, and they are essentially less than the characteristic radius of curvature for the band-contact line. In this case  practically all electron orbits in the Brillouin zone, which are intersections  of the Fermi surface with planes perpendicular to the magnetic field, have the elliptic shape and lie near the band-contact line. In other words, a small region in the Brillouin zone determines the local  electron energy spectrum in the magnetic field almost for any point of the line. This spectrum has the form \cite{m-sh18}:
 \begin{eqnarray}\label{39}
\varepsilon_{c,v}^l(p_3)&=&\varepsilon_{d}(p_3) \pm \!\left(\frac{e\hbar\alpha H|\cos\theta|}{c}l\right)^{1/2}\!, \\
\alpha&=&\alpha(p_3)=2(b_{11}b_{22})^{1/2}(1-\tilde a_{\perp}^2)^{3/2}, \label{40}
 \end{eqnarray}
where $l$ is a non-negative integer ($l=0$, $1$, \dots),
with the single Landau subband $l=0$ being shared between the branches ``$c$'' and ``$v$'', and $\theta=\theta(p_3)$ is the angle between the direction of the magnetic field and the tangent ${\bf t}={\bf t}(p_3)$ to the band-contact line at the point with a coordinate $p_3$, Fig.~\ref{fig5}. When $|\zeta- \varepsilon_d^0|$ and $\Delta$ are comparable, formula (\ref{39}) is valid in the leading order in the small parameter $(\Delta\cdot\tan\theta/LV)^2 \sim (\Delta\cdot\tan\theta/\varepsilon_0)^2$. It is clear that formula (\ref{39}) fails only for those points of the line for which $\theta$ is close to $\pi/2$. However, these points give a small contribution to the magnetization \cite{m-sh18}, and hence they do not introduce essential errors into results of its calculations.

Under the condition $\Delta\cdot\tan\theta/LV \ll 1$, spectrum (\ref{39}) can be rewritten in the form that is similar to Eq.~(\ref{7}),
\begin{equation}\label{41}
S(\varepsilon^{l},p_3)=\frac{S_0(\varepsilon^{l},p_3)}{|\cos\theta|} =\frac{2\pi\hbar e H}{c}l,
\end{equation}
where $S(\varepsilon^{l},p_3)$ is the area of the cross-section of the constant-energy surface $\varepsilon_{c,v}({\bf p})=\varepsilon^l$ by the plane perpendicular to the magnetic field  and passing through the point of the line with the coordinate $p_3$, and $S_0(\varepsilon^{l},p_3)$ is the area of the analogous  cross-section by the plane perpendicular to the line. Comparing Eq.~(\ref{41}) with the semiclassical quantization condition (\ref{12}) for spinless particles, one finds that $\gamma=0$. This result can be readily understood with Eq.~(\ref{13}). When the spin-orbit interaction is neglected, the orbital moment $L$ vanishes. On the other hand, if $\theta \neq \pi/2$, the electron orbits surround the nodal line, and the Berry phase $\Phi_B$ of the orbits  is equal to $\pi$ \cite{prl}. It is significant that this result for $\Phi_B$ remains true even if terms of higher orders in $p_1$ and $p_2$ are included in spectrum (\ref{38}). Besides, the spectrum in the magnetic field, Eq.~(\ref{41}), is not disturbed by the weak spin-orbit interaction when the gap $2\Delta_0(p_3)$ is induced along the nodal line. With this interaction, the semiclassical quantization condition (\ref{11}) comes into play. In this condition the $g$ factor comprises the two terms $g=g_1(p_3)+g_2(p_3)$ \cite{jetp}. The first term is determined by the Berry phase,
 \begin{eqnarray*}
g_1=2\frac{m}{m_*}\left (1-\frac{\Delta_0}{\sqrt{(\zeta-\varepsilon_d)^2-\Delta_{\rm min}^2({\bf n})+\Delta_0^2}}\right ) ,
 \end{eqnarray*}
while the second one is caused by the orbital moment $L$,
\begin{eqnarray*}
g_2=2\frac{m}{m_*}\frac{\Delta_0}{\sqrt{(\zeta-\varepsilon_d)^2-\Delta_{\rm min}^2({\bf n})+\Delta_0^2}} ,
 \end{eqnarray*}
where $m_*$ is the cyclotron mass of the electron orbit in the magnetic field directed along the unit vector ${\bf n}$, and $2\Delta_{\rm min}({\bf n})$ is the minimal indirect gap between the bands ``$c$'' and ``$v$'' in the plane of this orbit (it follows from Eqs.~(\ref{38}) that $\Delta_{\rm min}({\bf n})\approx \Delta_0(1-\tilde a_{\perp}^2)^{1/2}$ at $\Delta\cdot\tan\theta/LV \ll 1$). The formulas for $g_1$ and $g_2$ show that the Berry-phase term dominates over the the orbital-moment term at $\Delta_0 \to 0$. However, in any case one has $g=g_1+g_2=2m/m_*$, and the quantization condition (\ref{11}) really reduces to Eq.~(\ref{41}).

\subsection{Magnetization} \label{sec4.2}

As in the case of the Weyl and Dirac semimetals, the total magnetic susceptibility of the nodal-line semimetals consists of its special part determined by the electron states located in the vicinity of the band-contact line and a practically constant  background term specified by electron states located far away from this line. It is the special part that is responsible for dependences of the susceptibility on the magnetic field, temperature, and the chemical potential $\zeta$ when this $\zeta$ lies inside or close to the narrow energy interval from $\varepsilon_d^0- \Delta$ to $\varepsilon_d^0 +\Delta$. Below we consider only the special contributions to the magnetization and to the magnetic susceptibility.

The magnetic-field behavior of the susceptibility essentially depends on interrelation between the four parameters: the width $2\Delta$ of the degeneracy-energy dispersion along the nodal line, the temperature  $T$, the characteristic spacing  $\Delta\varepsilon_H=\hbar\omega_c$ between the Landau subbands, and the chemical potential $\zeta$ measured relative to the critical energy $\varepsilon_{\rm min}$ or $\varepsilon_{\rm max}$ of the line. In weak magnetic fields $H\ll H_T$, when $\Delta\varepsilon_H\ll T$, the special part of the magnetic susceptibility is practically  independent of $H$, whereas at $H>H_T$ the magnetization becomes a nonlinear function of $H$. The boundary $H_T$ between the regions of weak and strong magnetic fields, $\Delta\varepsilon_H(H_T) \sim T$, is given by the same estimate as in the case of the Dirac (Weyl) semimetals.

The giant anomaly in the magnetic susceptibility of metals with band-contact lines was theoretically studied both for weak \cite{m-sv} and for strong \cite{m-sh} magnetic fields. This anomaly is determined by the electron states located near the points of the $3\frac{1}{2}$-order electron topological transitions occurring at $\zeta=\varepsilon_{\rm min}$ and $\zeta=\varepsilon_{\rm max}$.
In fact, it was assumed in the papers \cite{m-sv,m-sh} that the parameter $\Delta$ is large, and  it exceeds the other three parameters, i.e.,  $\Delta \gg T$, $\Delta\varepsilon_H$, $\zeta-\varepsilon_{\rm min}$ (or $\zeta-\varepsilon_{\rm max}$). In this situation each critical point can be considered independently. On the other hand, the magnetic susceptibility for a nodal-line semimetal with a band-contact ring characterized by constant $\varepsilon_d$ (i.e. by $\Delta=0$) was estimated by Koshino and Hizbullah \cite{kosh15} in the case of the weak magnetic fields. The ring was described by the model proposed in Ref.~\cite{balents}. Using a similar model for the nodal ring with $\Delta=0$, the quantum oscillations of the density of the electron states, the susceptibility, and the resistivity were theoretically studied in Refs.~\cite{liu17,pal16,li-prl18,yang18,orosz}. The magnetization of the semimetals with the band-contact lines of arbitrary shapes and at small but  nonzero $\Delta$ was calculated in Refs.~\cite{m-sh16,m-sh18} for the weak and strong magnetic fields, including the region of the de Haas - van Alphen oscillations. These oscillations were further analyzed in Ref.~\cite{fnt18}.

At $T=0$ the general formula for the magnetization associated with a band-contact line in a nodal-line semimetal has the form  \cite{m-sh18}:
 \begin{eqnarray}\label{42}
  {\bf M}(\zeta,H)\!=\!\frac{e^{3/2}H^{1/2}} {2\pi^2\hbar^{3/2}c^{3/2}}\!\!\int_{0}^{L}\!\!\!\!\!dp_3 |\cos\theta|^{1/2}\nu\!\sqrt{\alpha(p_3)} K(u){\bf t},
  \end{eqnarray}
where the integration in the Brillouin zone is carried out over the  length $L$ of the line; $\theta=\theta(p_3)$ is angle between ${\bf H}$ and the unit vector ${\bf t}={\bf t}(p_3)$ tangent to the line at the point $p_3$; $\nu=\nu(p_3)$ is a sign of $\cos\theta$;
 \begin{eqnarray}\label{43}
 K(u)\!\!\!&=&\!\!\! \frac{3}{2}\zeta(-\frac{1}{2},\![u]\!+\!1\!)+\sqrt{u}([u]+\frac{1}{2}), \end{eqnarray}
$\zeta(s,a)$ is the Hurwitz zeta function, the quantity $u$ is similar to that defined by Eq.~(\ref{23}),
 \begin{equation}\label{44}
 u=\frac{[\zeta-\varepsilon_d(p_3)]^2 c}{e\hbar  \alpha(p_3) H|\cos\theta|}=\frac{cS(p_3)}{2\pi e\hbar H},
 \end{equation}
$S(p_3)$ is the area of the Fermi-surface cross section
by the plane perpendicular to the magnetic field and passing through the point of the line with the coordinate $p_3$, $[u]$ is the integer part of $u$. In Eq.~(\ref{42}), the two-fold degeneracy of the conduction and valence bands in spin is assumed.
In the case of a noncentrosymmetric semimetal with a strong spin-orbit interaction (e.g.,  PbTaSe$_2$ \cite{bian}), where this degeneracy is absent, the right hand side of formula (\ref{42})  should be divided by two. In obtaining formula (\ref{42}), it was also supposed that Eqs.~(\ref{39}) and (\ref{40}) are valid at all angles $\theta$ including $\theta=\pi/2$. As was mentioned above, this supposition does not lead to an essential error in the magnetization since the parts of line where $\theta \sim \pi/2$ give a small contribution to expression (\ref{42}).

For nonzero $T$, the magnetization $M_i(\zeta,H,T)$ can be calculated with the relationship:
 \begin{eqnarray}\label{45}
 M_{i}(\zeta,H,T)=-\int_{\!-\infty}^{\infty}\!\! d\varepsilon M_{i}(\varepsilon,H,0)f'(\varepsilon),
 \end{eqnarray}
where $f'(\varepsilon)$ is the derivative of the Fermi function,
\begin{equation}\label{46}
 f'(\varepsilon)=-\left[4T\cosh^2\left(\frac{\varepsilon- \zeta}{2T}\right)\right]^{-1}.
 \end{equation}

In the topological semimetals, charge carriers (electron and holes) are located near the band-contact line, and their chemical potential $\zeta$ generally depends on the magnetic field, $\zeta=\zeta(H)$. This dependence can be derived from the condition that the charge carrier density $n$ does not vary with increasing $H$,
\begin{equation}\label{47}
n(\zeta,H)=n_0(\zeta_0),
\end{equation}
where $n_0$ and $\zeta_0$ are the density and the chemical potential at $H=0$. On calculating $\zeta(H)$, one can find the magnetization as a function of $n_0$ or $\zeta_0$, inserting $\zeta(H)$ into Eq.~(\ref{42}). As in the case of the Dirac and Weyl semimetals, if there are no other electron pockets stabilizing the chemical potential in the nodal-line semimetal, the dependence $\zeta(H)$ has a noticeable effect on the magnetization in the ultra-quantum regime or when $H$ approaches its boundary $H_1$. However a specific feature of certain nodal-line semimetals is that the spacing $\Delta \varepsilon_H$ between the Landau subbands may become larger than their width even at $H \ll H_1$. In this situation the spectrum (\ref{39}) transforms, in fact, into the spectrum of a two-dimensional electron system since different Landau subbands $\varepsilon_{c,v}^l(p_3)$ do not overlap, and they look like broadened Landau levels. In this quasi-two-dimensional case the chemical potential essentially depends on the magnetic field \cite{Sh}, and its dependence noticeably changes the shape of the de Haas - van Alphen oscillations; see below.

Formulas (\ref{42}), (\ref{45}) (and the appropriate expression for the $\Omega$ potential) enable one to analyze the cases of the weak and strong magnetic fields, including the region of the de Haas- van Alphen oscillations and the ultra-quantum regime \cite{m-sh18}. We now consider these cases in more detail.

\subsubsection{Weak magnetic fields}

\begin{figure}[tbp] 
 \centering  \vspace{+9 pt}
\includegraphics[scale=.99]{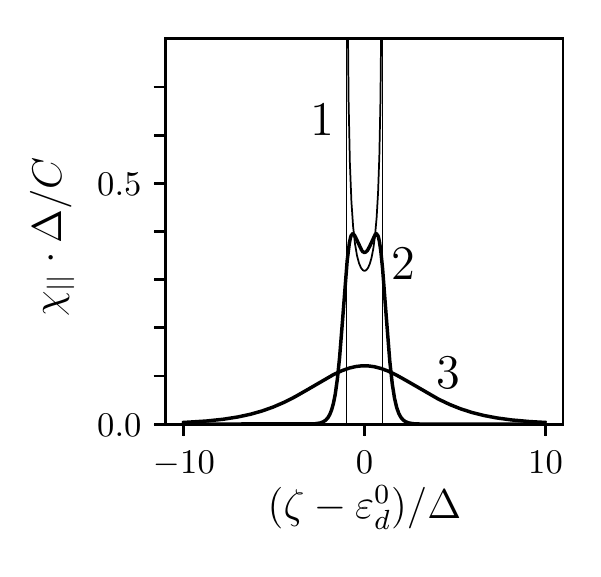}
\caption{\label{fig7} The dependence of $\chi_{\parallel}$ on the chemical potential $\zeta$ at 1) $T\to 0$, Eq.~(\ref{51}) 2) $T/\Delta=0.25$ and 3) $T/\Delta=2$, Eq.~(\ref{49}). The susceptibility is measured in units of $C/\Delta$ where the constant $C$ is defined by formula (\ref{52}).
 } \end{figure}

In the weak magnetic fields $H\ll H_T$, when $\Delta\varepsilon_H\ll T$, one has the following expressions for the magnetization and the longitudinal magnetic susceptibility $\chi_{\parallel}$ defining the magnetization component $M_{\parallel}=\chi_{\parallel}H$ parallel to the magnetic field \cite{m-sh16,m-sh18}:
\begin{eqnarray}\label{48}
{\bf M}(\zeta,H,T)\!&=&\!\frac{e^{2}H} {12\pi^2\hbar c^{2}}\!\!\int_{0}^{L}\!\!\!\!\!dp_3 \alpha(p_3)f'(\varepsilon_d)\cos\theta\,{\bf t}. \\
   \chi_{\parallel}\!&=&\!\frac{e^2}{12\pi^2\!\hbar c^2}\!
 \!\!\int_{0}^{L}\!\!\!\!\!dp_3\alpha(p_3)f'(\varepsilon_d)
 \cos^2\!\theta, \label{49}
 \end{eqnarray}
where $f'(\varepsilon_d)$ and $\alpha(p_3)$ are given by Eqs.~(\ref{46}) and (\ref{40}), respectively. At low temperatures  $T\ll 2\Delta$,  formula (\ref{49}) yields  $\chi_{\parallel}(\zeta) =0$ if $\zeta$ does not lie between $\varepsilon_{\rm min}=\varepsilon_d^0-\Delta$ and  $\varepsilon_{\rm max}=\varepsilon_d^0+\Delta$. In order to gain some insight into the behavior of $\chi_{\parallel}(\zeta)$ in the  interval $|\zeta-\varepsilon_d^0|<\Delta$, consider a simple model of the nodal line. Let the band-contact line be a circle with constant $b_{11}$, $b_{22}$, $\tilde a_{\perp}^2$, and with
   \begin{equation}\label{50}
 \varepsilon_d(p_3)=\varepsilon_d^0+\Delta \cos\!\left(\!\!\frac{2\pi p_3n}{L}\!\right),
 \end{equation}
where $n$ is an integer ($n=6$ in Fig.~\ref{fig5}). Within this model that may be appropriate for Ca$_3$P$_2$ \cite{xie}, it follows from Eq.~(\ref{49}) that \cite{m-sh16}:
\begin{eqnarray}\label{51}
 \chi_{\parallel}(\zeta)\!=\!\frac{C} {\pi \sqrt{\Delta^2-(\zeta-\varepsilon_d^0)^2}}=\frac{C} {\pi \sqrt{(\varepsilon_{\rm max}-\zeta)(\zeta-\varepsilon_{\rm min})}},
 \end{eqnarray}
where
\begin{eqnarray}\label{52}
 C=-\frac{e^2}{12\pi^2\!\hbar c^2}
L(b_{11}b_{22})^{1\!/2}\!(1-\tilde a_{\perp}^2)^{3\!/2}\!\cos^2\theta_0 ,
 \end{eqnarray}
and $\theta_0$ is the angle between the magnetic field and the plane of the nodal line. When $\zeta$ tends, e.g., to  $\varepsilon_{\rm max}= \varepsilon_d^0 +\Delta$, the giant anomaly in the susceptibility, $\chi_{\parallel}\propto (\varepsilon_{\rm max}- \zeta)^{-1/2}$, occurs in agreement with Ref.~\cite{m-sv}. Of course, at $|\varepsilon_{\rm max} -\zeta|\lesssim T$ the divergence of $\chi_{\parallel}$ in Eq.~(\ref{51}) is truncated as in the case of the Dirac points. In the middle of the interval $\varepsilon_{\rm min}\le \zeta \le \varepsilon_{\rm max}$ the susceptibility is still large due to a small value of $\Delta$ in the denominator of Eq.~(\ref{51}), and $\chi_{\parallel}$ essentially depends on $\zeta$, Fig.~\ref{fig7}.

When the temperature is not small as compared to $\Delta$ (and $T\gg \Delta\varepsilon_H$), the $\zeta$-dependence of  $\chi_{\parallel}(\zeta,T)$ is shown in Fig.~\ref{fig7}. In the limiting case when $\Delta$ is so small (or the temperature is so high) that $\Delta\ll T$ (the interrelation between $\Delta$ and $\Delta\varepsilon_H$ may be arbitrary), the explicit formula for $\chi_{\parallel}(\zeta,T)$ can be obtained \cite{m-sh16}:
\begin{eqnarray}\label{53}
 \chi_{\parallel}(\zeta,T)\!=\!\frac{C}{4T\cosh^2[(\varepsilon_d^0- \zeta)/2T]}.
 \end{eqnarray}
This formula, in fact, reproduces the result of Koshino and Hizbullah \cite{kosh15}, and it well describes $\chi_{\parallel}(\zeta,T)$ calculated with Eqs.~(\ref{49}) even at $T/\Delta \gtrsim 2$.

When the angle $\theta_0$ differs from zero and $\pi/2$, there is also a nonzero component $M_{\perp}$ of the magnetization that is perpendicular to the magnetic field ${\bf H}$. This component lies in the plane passing through the vector ${\bf H}$ and the normal to the plane of the nodal line, and it can be calculated with Eq.~(\ref{48}). It turns out that the magnetic susceptibility $\chi_{\perp}$ defined by the relation $M_{\perp}=\chi_{\perp}H$ is described by formulas (\ref{51})-- (\ref{53}) in which $\cos^2\theta_0$ should be replaced by $\cos\theta_0\sin\theta_0$ \cite{m-sh16}. This $\chi_{\perp}$ determines the magnetic torque  $K_{\phi}=\chi_{\perp}H^2$.

Koshino and Hizbullah \cite{kosh15} estimated the spin $\chi_{\rm s}$ and cross $\chi_{\rm s-o}$ susceptibilities for the nodal ring. These susceptibilities are similar to those considered in Sec.~\ref{sec3.1} for the Dirac points. It follows from their results that $\chi_{\rm s}$ and $\chi_{\rm s-o}$  are relatively small as compared to the orbital part of the susceptibility considered above if $\Delta$ and $|\zeta-\varepsilon_d^0|$ are essentially less than $mV^2$ where $V^2\sim (b_{11}b_{22})^{1/2}$.

Consider now an example of the band-contact line terminating on the opposite faces of the Brillouin zone. In particular, this situation occurs in BiTeI and BiTeCl although these crystals hardly be assigned to the representative set of the nodal-line semimetals since the parameter $\Delta$ is sufficiently large in them ($2\Delta\sim 0.3$ eV in BiTeI \cite{bahramy11}). It is well known that a large Rashba-type spin splitting induced by the strong spin-orbit interaction occurs in the bulk bands of the crystals BiTeI \cite{bahramy11,mura13,vangen,Ye,park15} and BiTeCl \cite{chen14,martin14,xiang-prb15} which have no center of the inversion. The splitting of the conduction band $\varepsilon_c({\bf p})$ into the two spin subbands $\varepsilon_c^{\pm}({\bf p})$  in the vicinity of the $A$-$\Gamma$-$A$ axis (the $z$ axis) of the Brillouin zone can be described by the following Hamiltonian:
\begin{eqnarray}\label{54}
 \hat H=\left[\varepsilon_d(p_z)+ \frac{p_x^2+p_y^2}{2m_{\parallel}}\right]\sigma_0+ v_R\,{\bf \sigma}\cdot [{\bf e}_z \times {\bf p}],
 \end{eqnarray}
which leads to the dispersion relation:
 \begin{eqnarray}\label{55}
\varepsilon_c^{\pm}({\bf p})= \varepsilon_d(p_z)+ \frac{p_x^2+p_y^2}{2m_{\parallel}} \pm v_R\sqrt{p_x^2+p_y^2}\,.
 \end{eqnarray}
Here the unit vector ${\bf e}_z$ is directed along the $z$ axis,  $v_R=v_R(p_z)$ and $m_{\parallel}=m_{\parallel}(p_z)$ are the Rashba parameters, $\sigma_0$ and ${\bf \sigma}$ are the unit and Pauli matrices. It follows from Eq.~(\ref{55}) that at a fixed $p_z$ the minimum of the conduction band, $\varepsilon_{c,{\rm min}}^{-}(p_z)$, occurs at the distance $p_0=v_R m_{\parallel}$ from the $z$ axis and lies below $\varepsilon_d(p_z)$, $\varepsilon_{c, {\rm min}}^{-}(p_z) =\varepsilon_d(p_z)- p_0^2/2m_{\parallel}$.  It is also clear that there is no spin splitting of the conduction band when ${\bf p}$ lies in this axis ($p_x=p_y=0$) although the point $A$ in BiTeI and the point $\Gamma$ in BiTeCl are frequently named as the Dirac points. In reality, the $A$-$\Gamma$-$A$ axis is the band-contact line in which the two spin subbands of the conduction band merge. In BiTeI the points $A$ and $\Gamma$ correspond to $\varepsilon_{\rm min}$ and $\varepsilon_{\rm max}$ of $\varepsilon_d(p_z)$, respectively, whereas in BiTeCl the opposite correspondence takes place: $\varepsilon_{\rm min}= \varepsilon_d(\Gamma)$ and $\varepsilon_{\rm max}=\varepsilon_d(A)$. The crystals BiTeI and BiTeCl are always doped, and the chemical potential $\zeta$ lies near $\varepsilon_{\rm min}$. With changing the doping, $\zeta$ crosses $\varepsilon_{\rm min}$, and the $3\frac{1}{2}$ order topological  transition occurs \cite{m-sh14,m-sh-jltp} which is known for these crystals as the transition from the ring torus to the so-called spindle torus \cite{Ye,xiang-prb15}.

Dependences of the magnetic susceptibility of BiTeI on $T$ and $\zeta$ were measured by Schober et al. \cite{schober} in the region of weak magnetic fields, and it was found that there is a diamagnetic anomaly in $\chi_{zz}(\zeta)$ superimposed on a paramagnetic background when $\zeta$ crosses $\varepsilon_{\rm min}$. This anomaly for the dispersion relation (\ref{55})  was first studied theoretically by Boiko and Rashba \cite{boiko}.
Their results can be readily reproduced with the general formula (\ref{49}) if we take into account that $\alpha(p_3)=2v_R^2$ in the case of Eq.~(\ref{55}) and approximate $\varepsilon_d(p_z)$ near $\varepsilon_{\rm min}$ by the dependence $\varepsilon_d(p_z) \approx  \varepsilon_{\rm min} +p_z^2/2m_z$ where $m_z$ is an effective mass. At low temperatures and at $\zeta\ge \varepsilon_{\rm min}$, the contribution of the band-contact line to $\chi_{zz}$ has the form:
\begin{eqnarray}\label{56}
   \chi_{zz}=-\frac{e^2v_R^2\sqrt{2m_z}}{12\pi^2\!\hbar c^2\sqrt{\zeta-\varepsilon_{\rm min}}} ,
 \end{eqnarray}
where the additional factor $1/2$ has been introduced since the bands are not degenerate in spin. The other components of the susceptibility tensor and $\chi_{zz}$ at $\zeta< \varepsilon_{\rm min}$ are equal to zero. Temperature dependences of the susceptibility can be easily obtained either with  Eq.~(\ref{49}) or with Eqs.~(\ref{45}) and (\ref{56}). In particular, the magnitude of the diamagnetic anomaly is proportional to $T^{-1/2}$.

\subsubsection{Oscillations}

Consider now the case of sufficiently strong magnetic fields, $H>H_T$, and begin with the situation when the chemical potential does not lie near $\varepsilon_{\rm min}$ or $\varepsilon_{\rm max}$ of the nodal line, i.e., when $|\zeta -\varepsilon_{\rm min}| \sim |\zeta -\varepsilon_{\rm max}|$. In this case it follows from Eq.~(\ref{42}) \cite{m-sh18} that in the region of the magnetic fields where $T\ll \Delta\varepsilon_H \ll |\zeta-\varepsilon_{min}|$, $|\zeta-\varepsilon_{max}|$, $2\Delta$, the magnetization is described by the usual formula for the de Haas - van Alphen effect in three-dimensional metals \cite{Sh}, with the phase of the oscillations being shifted by a half of the period.  In particular, the $n$-th harmonic $\Delta M_n$ of the magnetization component $M_{\parallel}$ parallel to the magnetic field looks like:
\begin{eqnarray} \label{57}
 \Delta M_n \propto \frac{H^{1/2}}{n^{3/2}}\sin\!\left(\!\!2\pi n \!\left[\!\frac{F}{H}-\gamma \right] \pm \frac{\pi}{4}\right),
 \end{eqnarray}
where $F=S_{\rm ex}c/2\pi\hbar e$ is the frequency determined by the extremal (in $p_3$) cross-section area $S_{\rm ex}$ of the Fermi surface, $\gamma=1/2 -\Phi_B/2\pi$ \cite{prl}, and the offsets  $\pm \pi/4$ refer to the minimal and maximal $S_{\rm ex}$, respectively. The phase shift of the oscillations is due to the zero value of $\gamma$ that is caused by the Berry phase $\pi$ for the electron orbits surrounding this line, see Sec.~\ref{sec4.1}. This shift is the characteristic feature of such orbits in crystals with band-contact lines \cite{prl}. As in usual metals \cite{Sh}, the dependence $\zeta(H)$ is sufficiently weak in this region of the magnetic fields and practically has no effect on the oscillations.

\begin{figure}[tbp] 
 \centering  \vspace{+9 pt}
\includegraphics[scale=0.90]{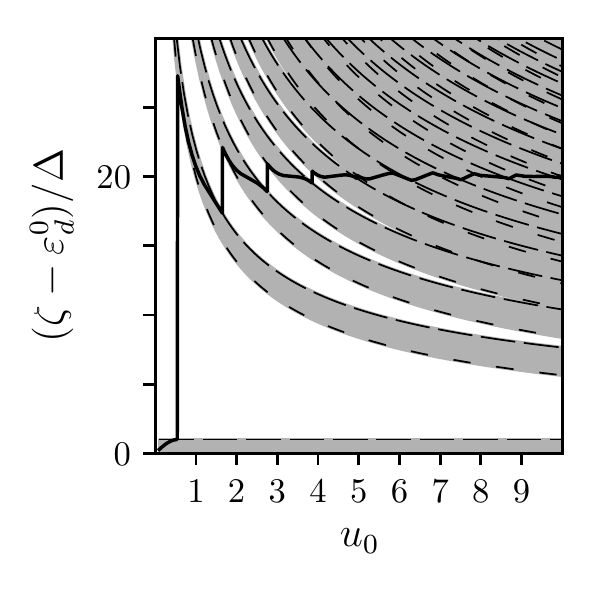}
\caption{\label{fig8} The dependence of chemical potential $\zeta$ on $1/H$ calculated with Eq.~(\ref{47}) for rhombohedral graphite at $\zeta(H=0)\equiv \zeta_0=\varepsilon_d^0+20\Delta$. The magnetic field is directed along the $z$ axis (the axis of the band-contact helix); $u_0\equiv (\zeta_0-\varepsilon_d^0)^2 c/e\hbar \alpha  H\cos\theta$,  $\varepsilon_d(p_3)=\varepsilon_d^0 -\Delta \cos(6\pi p_3/L)$, $\cos\theta = 0.98$, $\alpha(p_3)=$const \cite{m-sh18}. We also mark the Landau subbands, Eq.~(\ref{39}), by the dark background, and the short and long dashes indicate the lower and the upper edges of these subbands, respectively. The crossover described in the text occurs at $u_0 \sim 5$.
 } \end{figure}

In the doped nodal-line semimetals the additional relation $2\Delta \ll |\zeta-\varepsilon_{min}|$, $|\zeta-\varepsilon_{max}|$ is fulfilled. In this case, with increasing magnetic field when $T< 2\Delta \ll \Delta\varepsilon_H \ll |\zeta-\varepsilon_d^0|$, the three-dimensional spectrum (\ref{39}) transforms, in fact, into the spectrum of a two-dimensional electron system if the width of the Landau subbands $\varepsilon_{c,v}^l(p_3)$ is of the order of $2\Delta$. In this situation different Landau subbands do not overlap, and they look like broadened Landau levels. In this magnetic-field region the $n$-th harmonic $\Delta M_n$ of the longitudinal magnetization $M_{\parallel}$ takes the form characteristic of two-dimensional metals \cite{luk04}:
\begin{eqnarray} \label{58}
 \Delta M_n \propto \frac{1}{n}\sin\!\left(\!\!2\pi n \!\left[\!\frac{F_{2D}}{H}-\gamma \right]\right),
 \end{eqnarray}
where $F_{2D}=(S_{\rm max}+S_{\rm min})c/4 \pi \hbar e$, $S_{\rm max}$ and $S_{\rm min}$ are the maximum and minimum cross-section areas of the Fermi surface, and  $\gamma=0$ \cite{m-sh18}. When $H$ changes within  this region of the magnetic fields, the chemical potential $\zeta(H)$ moves together with one of these levels, and then, at a certain value of $H$, it jumps from this level to the neighboring one \cite{Sh}. The crossover from the  three-dimensional electron spectrum to the quasi-two-dimensional one occurs at the field $H_{cr}$ determined by the condition $\Delta \varepsilon_H(H_{cr}) \sim \Delta$, while the boundary $H_1$ of the ultra-quantum regime is essentially larger than  $H_{cr}$, and it is found from $\Delta\varepsilon_H(H_1) \sim |\zeta-\varepsilon_d^0|$ in this quasi-two-dimensional case. Using Eq.~(\ref{39}), one obtains the following estimates for $H_1$ and $H_{cr}$:
\begin{eqnarray} \label{59}
H_1&\sim& F_{2D}\sim \frac{(\zeta-\varepsilon_d^0)^2 c}{e\hbar \alpha },  \\
H_{cr}&\sim& \frac{(S_{\rm max}-S_{\rm min}) c}{2\pi e\hbar  }\sim \frac{4\Delta |\zeta-\varepsilon_d^0| c}{e\hbar \alpha }\sim \frac{4\Delta}{|\zeta-\varepsilon_d^0|}H_1. \nonumber
 \end{eqnarray}
An example of the dependence $\zeta(H)$ both at $H<H_{cr}$ and $H>H_{cr}$ is shown in Fig.~\ref{fig8}.

Strictly speaking, the quasi-two-dimensional electron spectrum in magnetic fields does not develop for all electrons in a topological semimetals with small $\Delta$ since if $\cos\theta$ tends to zero in  some part of the line, the $\Delta\varepsilon_H$ becomes less than $2\Delta$ there; see Eq.~(\ref{39}). For the quasi-two-dimensional spectrum to manifest itself, a change of the quantity $u\propto 1/\cos\theta$ defined by  Eq.~(\ref{44}) has to be less than unity along an essential portion of the line, i.e., $\cos\theta$ may change only within a sufficiently small interval in this portion. This imposes a restriction on the shape of the nodal line. It is clear that the spectrum of this kind can appear for a straight band-contact line, i.e., for a symmetry axis, since $\theta(p_3)$ is constant in this case. The quasi-two-dimensional spectrum is also possible in the case of  band-contact lines terminating on the opposite faces of the Brillouin zone for a certain region of the magnetic-field directions. In particular, this situation takes place in rhombohedral graphite in which the nodal line has the shape of a helix \cite{mcclure}, and the quasi-two-dimensional spectrum is realized when the magnetic field $H>H_{cr}$ is perpendicular to the basal plane of the crystal  \cite{m-sh18}. This type of the spectrum can also occur for a closed band-contact line composed of nearly straight arcs. This situation appears to take place in ZrSiS \cite{schoop,neupane1}.

\begin{figure}[tbp] 
 \centering  \vspace{+9 pt}
\includegraphics[scale=.90]{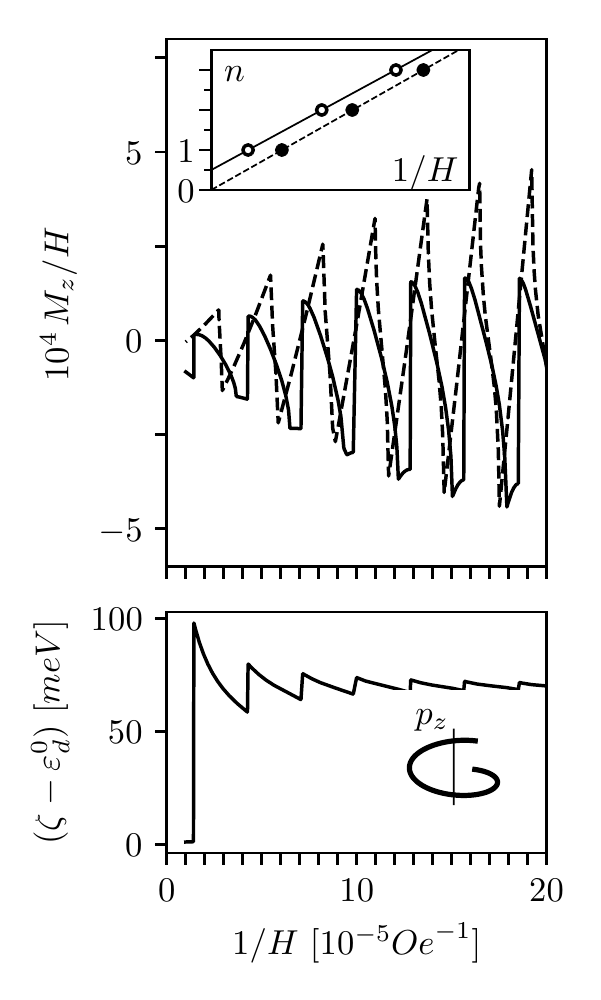}
\caption{\label{fig9}
Top: Dependences of $M_z/H$ on $1/H$ for rhombohedral graphite at $T=0$ and the magnetic field directed along the $z$ axis. The dependences are  calculated numerically with Eqs.~(\ref{42})-- (\ref{44}); $\alpha(p_3)$, $\cos\theta$, are $\varepsilon_d(p_3)$ are the same as in Fig.~\ref{fig8}; $\Delta=1$ meV \cite{m-sh18}. The dashed line corresponds to the constant chemical potential $\zeta-\varepsilon_d^0=70$ meV, the solid line shows $M_z/H$ at $\zeta_0-\varepsilon_d^0=70$ meV, taking into account the $H$-dependence of $\zeta$ presented in the bottom panel. The inset: The Landau-level fan diagrams plotted with the positions of the maxima of the dashed and solid curves of the main panel. The diagrams give $\gamma=0$ and $\gamma=1/2$. Bottom: The $H$-dependence of the chemical potential calculated with Eq.~(\ref{47}) at $\zeta_0- \varepsilon_d^0 =70$ meV. The inset schematically shows the band-contact helix in rhombohedral graphite.
 } \end{figure}

The strong dependence $\zeta(H)$ for the quasi-two dimensional spectrum noticeably changes the shape of the de Haas - van Alphen oscillations and can mask the correct value of the Berry phase $\Phi_B$ when $\Phi_B$ is measured with these oscillations \cite{m-sh18,fnt18}. In particular, if the quasi-two dimensional spectrum occurs for all the electrons near the nodal line, one can find  $\gamma=1/2$ from these oscillations although the Berry phase is still equal to $\pi$, and the true $\gamma=0$. To illustrate this statement, in Fig.~\ref{fig9} we present the $H$-dependence of $M_z/H$ for rhombohedral graphite when the magnetic field is perpendicular to the basal plane of the crystal (i.e., $H$ is parallel to the $z$ axis). The de Haas - van Alphen oscillations are clearly visible in the figure. The dashed line shows the oscillations calculated at constant $\zeta$. As expected,  the Landau-level fan diagram shown in  the inset yields $\gamma=0$ for these oscillations. However, the electron spectrum is quasi-two-dimensional for the whole interval of the magnetic fields  in Fig.~\ref{fig9}, the chemical potential and the magnetization $M_z$ exhibit jumps, and the positions of these jumps do not coincide with the positions of the sharp peaks in $M_z/H$ calculated at a constant chemical potential. In other words, the dependence $\zeta(H)$ leads to the shift of the oscillations, and this shift imitates a change of $\gamma$. In particular, the appropriate Landau-level fan diagram suggests that $\gamma=1/2$. Interestingly, intermediate values of $\gamma$ (different from $0$ and $1/2$) can be found if there is an addition group of charge carriers in a nodal-line semimetal, i.e., if the electron groups with the quasi-two-dimensional and three-dimensional spectra coexist in this semimetal \cite{fnt18}.

Recently, the de Haas - van Alphen \cite{Hu,kumar,Hu2,Hu1,Hu18}, Shubnikov - de Haas \cite{Ali1,wang1,singha,pez,delft,guo19}, and thermoelectric power \cite{matus} oscillations in magnetic fields were experimentally investigated in ZrSiS family of the nodal-node semimetals, and intermediate values of the Berry phase (other than $0$ and $\pi$) were obtained for a number of the electron orbits. Taking into account the above considerations, we hypothesized \cite{m-sh18,fnt18}  that the essential dependence of chemical potential on the magnetic field reveals itself in these measurements, and this dependence is probably due to the coexistence of dissimilar electron groups in these materials.
Under this assumption, the observed splitting of the oscillation frequencies with changing the magnetic-field direction \cite{kumar,Hu1,Ali1,singha,guo19} can be explained by the crossover from the frequency  $F_{2D}\propto (S_{\rm max}+S_{\rm min})/2$ characteristic of the quasi-two-dimensional spectrum to the frequencies $F_{\rm max}\propto S_{\rm max}$ and $F_{\rm min}\propto S_{\rm min}$ inherent in the three-dimensional case.
Another reason for the intermediate values of $\gamma$ may be due to  the magnetic breakdown \cite{Sh,azbel,alex1,alex2}. However, although the combination frequencies inhering in the magnetic breakdown  were really detected in ZrSiS \cite{pez}, ZrGeSe\cite{Hu2}, and HfSiS \cite{delft}, the intermediate values of $\gamma$ were not found for the appropriate  orbits. On the other hand, the combination frequencies corresponding to the orbits $\beta \pm \alpha$ in   Refs.~\cite{Hu2,delft} can appear when a noticeable dependence of $\zeta$ on $H$ occurs \cite{nakano}. As to the so-called magnetic interaction \cite{Sh} which can also generate the combination frequencies and change the phase of the oscillations, this interaction seems small in these semimetals, see the Supplemental Material to Ref.~\cite{delft}.

Finally, consider the case when the chemical potential is close to the energy $\varepsilon_{\rm min}$ or $\varepsilon_{\rm max}$. For definiteness, we imply that this energy is $\varepsilon_{\rm min}$, i.e.,  $|\zeta-\varepsilon_{\rm min}| \ll \Delta$. It is this case that is realized in BiTeI and BiTeCl. In this situation one has
 \begin{equation*}
H_1\sim \frac{(\zeta-\varepsilon_{\rm min})^2 c}{e\hbar \alpha},
 \end{equation*}
the oscillations at $H_T< H <H_1$ are always three-dimensional, and the $H$-dependence of the chemical potential is inessential except for the last oscillations at $H\sim H_1$. Hence, one may expect to find $\gamma=0$ for the electron orbits surrounding the nodal line. The experimental investigations of the Shubnikov - de Haas oscillations in BiTeI \cite{mura13,vangen,Ye,park15} and BiTeCl \cite{xiang-prb15} did reveal that $\gamma=0$ for the extremal orbits lying on the Fermi surfaces $\varepsilon_c^-({\bf p})=\zeta$ and $\varepsilon_c^+({\bf p})=\zeta$ defined by Eq.~(\ref{55}).

\subsubsection{The ultra-quantum regime}

\begin{figure}[tbp] 
 \centering  \vspace{+9 pt}
\includegraphics[scale=.90]{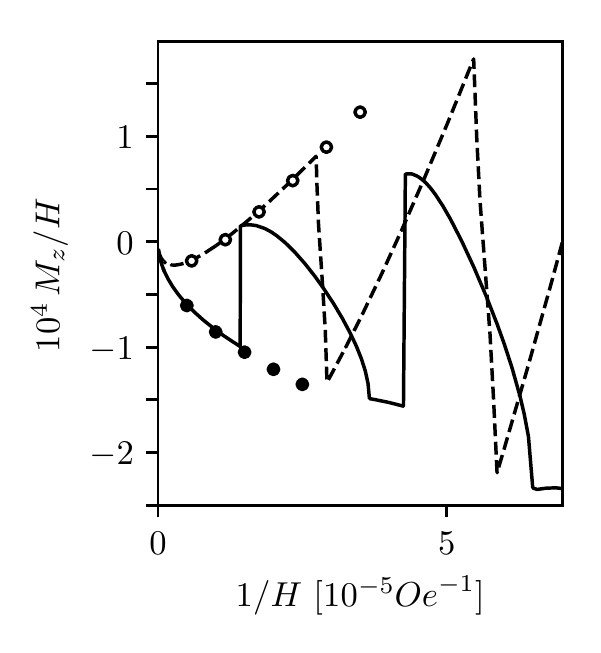}
\caption{\label{fig10}
Dependences of $M_z/H$ on $1/H$ for rhombohedral graphite at $T=0$ and the magnetic field directed along the $z$ axis. The dependences are  calculated numerically with Eqs.~(\ref{42})-- (\ref{44}); the parameters are the same as in Fig.~\ref{fig9}. The dashed line corresponds to the constant chemical potential $\zeta-\varepsilon_d^0=70$ meV, the solid line shows $M_z/H$ at $\zeta_0-\varepsilon_d^0=70$ meV, taking into account the $H$-dependence of $\zeta$ presented in the bottom panel of Fig.~\ref{fig9}. The field $H_1\sim 35$ kOe. The circles and the black dots show $M_z/H$ calculated with Eq.~(\ref{60}), but the dots corespond to the case when the second term in Eq.~(\ref{60}) is omitted.}
\end{figure}

In the ultra-quantum regime when $H>H_1$, the de Haas - van Alphen oscillations disappear. If the chemical potential is not close to the critical energies $\varepsilon_{\rm min}$ and $\varepsilon_{\rm max}$, the differences $ |\zeta-\varepsilon_{min}|\sim  |\zeta-\varepsilon_{max}|$ are of the order of $\Delta$ or even essentially exceed $\Delta$ for the doped semimetals. In this situation, at $H\gg H_1$ when  $|\zeta-\varepsilon_{min}|$, $|\zeta-\varepsilon_{max}|\ll \Delta\varepsilon_H$, the argument $u$ of the function $K(u)$ in Eq.~(\ref{42}) is small practically for all points of the band-contact line, and hence $K(u)= (3/2)\zeta(-1/2,1)+\sqrt{u}/2$. Then formula (\ref{42}) gives $M_i=c_1H^{1/2}+c_2$, with the coefficients $c_1$, $c_2$ depending on the direction of the magnetic field and on the shape of the band-contact line. In particular, we find the following expression for longitudinal component of the magnetization \cite{m-sh16}:
\begin{eqnarray} \label{60}
   M_{\parallel}(H)&\approx&\frac{3\zeta(-1/2,1)e^{3/2}H^{1/2}} {4\pi^2\hbar^{3/2}c^{3/2}}\!\!\int_{0}^{L}\!\!\!\!\!dp_3 |\cos\theta|^{3/2}\sqrt{\alpha(p_3)} \nonumber\\
   &+&\frac{e}{4\pi^2\hbar^2 c}\!\!\int_{0}^{L}\!\!\!\!\!dp_3 |[\zeta-\varepsilon_d(p_3)]\cos\theta|.
  \end{eqnarray}
As an example, the $H$-dependence of $M_z$ in the ultra-quantum regime for rhombohedral graphite is shown in Fig.~\ref{fig10}. It is seen that the dependence given by Eq.~(\ref{60}) well describes $M_z(H)$ at constant $\zeta$. However in the ultra-quantum regime the chemical potential tends to $\varepsilon_d^0$, Fig.~\ref{fig9}, and the second term in formula (\ref{60}) becomes small. For this reason,  $M_z(H)$ in the ultra-quantum regime is well described, in fact, by the first term of this formula when the $H$-dependence of $\zeta$ is taken into account. Note also that similar to the case of the Weyl (Dirac) semimetals, there is an analogy between the formulas in the ultra-quantum regime and the $2+1$ quantum electrodynamics with massless fermions \cite{niss18}.

If the chemical potential lies near the critical energy $\varepsilon_{\rm min}$ or $\varepsilon_{\rm max}$, at $H> H_1$ the de Haas - van Alphen oscillations disappear, but the spacing $\Delta \varepsilon_H$ between the Landau subbands is much less than  $2\Delta$. In this case one arrives at  $M_{\parallel} \propto H^{3/4}$ \cite{m-sh,m-sh16}. If $H$ further increases, and $\Delta \varepsilon_H$ exceeds $2\Delta$, the dependence $M_{\parallel}\propto H^{3/4}$ crosses over to $M_{\parallel}\propto H^{1/2}$.

The dependence $M_{\parallel}\propto H^{3/4}$ has to occur for BiTeI and BiTeCl in the ultra-quantum regime since $2\Delta$ is sufficiently large for these materials, and $|\zeta-\varepsilon_{\rm min}|\ll 2\Delta$. In particular, formula (\ref{42}) in the case of the spectrum (\ref{55}) leads to the following expression for longitudinal magnetization \cite{m-sh16}:
\begin{eqnarray}\label{61}
 M_{\parallel}(H)\!=\!-\frac{0.158e^{7/4}(2v_R)^{3/4}\sqrt{2m_z}} {2\pi^2\hbar^{5/4}c^{7/4}}\cdot H^{3/4}(\cos\theta)^{7/4},
  \end{eqnarray}
where the additional factor $1/2$ has been introduced analogously to  Eq.~(\ref{56}), the effective mass $m_z$ is determined by the expansion $\varepsilon_d(p_z)\approx \varepsilon_{\rm min}+p_z^2/2m_z$, and  $\theta$ is the angle between the magnetic field and the $z$ axis.

As in the case of the weak magnetic fields, the magnetization in the region of strong magnetic fields has the component ${\bf M}_{\perp}$ that is perpendicular to ${\bf H}$ \cite{m-sh16}.

\section{Conclusions}\label{conc}

In this review article we show that by now the theory of the magnetic susceptibility for the Weyl, Dirac, and nodal-line  semimetals has already been developed adequately, and we present the appropriate results for the susceptibility in the weak and strong magnetic fields, including the region of the de Haas - van Alphen oscillations and the ultra-quantum regime. The theory takes into account the tilt of the Dirac and Weyl spectra and the dispersion of the degeneracy energy along the band-contact line in the nodal-line semimetals. The following characteristic features of the magnetic susceptibility are analyzed in the article: The strong specific dependences of the susceptibility on the chemical potential and temperature for the weak magnetic fields, the phase shift of the de Haas - van Alphen oscillations, and the specific magnetic-field dependences of the magnetization in the ultra-quantum regime. These features can serve as hallmarks of the electron spectra inhering in the topological semimetals. The presented results are illustrated, discussing the following Dirac, Weyl, and nodal-line semimetals: Na$_3$Bi, Cd$_3$As$_2$, TaAs family, Ca$_3$P$_2$, ZrSiS, rhombohedral graphite, and also the bulk Rashba semiconductors BiTeI and BiTeCl in which a band-contact line exists. As to experimental investigations, there is a large number of experimental works that explore one of these hallmarks, viz., the phase shift of the oscillating part of the magnetization (or of the conductivity) by a half of the period. This offset is generally considered as evidence of a Weyl (Dirac) point or a nodal-line in a material under study. Interestingly, the phase shifts different from the predicted one were found in a number of these experiments. Possible explanations of such shifts are discussed in the article, too. As to other characteristic features of the magnetic susceptibility, there are only several experimental works that deal with non-oscillating part of the magnetization of the Weyl semimetals. In this context, extended experimental studies of the magnetization could provide additional useful information on the topological semimetals.

\end{document}